\documentclass[aip,jcp,reprint,amsbsy,amsfont,amsmath,floatfix,floats,
                          showpacs,showkeys,longbibliography]{revtex4-1}

\usepackage{dcolumn}
\usepackage{bm}
\usepackage{graphicx}
\usepackage{color}

\newcommand*\Rt[1]{$\sqrt{#1}$}

\begin{document}

\title{Simulations of submonolayer Xe on Pt$(111)$: \\
                 the case for a chaotic low temperature phase}
%                        thermodynamic and structural analyses}

\author{Anthony D. Novaco}\email[E-mail: ]{novacoad@lafayette.edu}
\affiliation{Department of Physics, Lafayette College \\ Easton,
                                                   Pennsylvania 18042, USA}

\author{Jessica Bavaresco}\email[E-mail: ]{jessica.bavaresco@oeaw.ac.at}
\affiliation{Institute for Quantum Optics and Quantum Information (IQOQI) \\
Austrian Academy of Sciences, Boltzmanngasse 3, A-1090 Vienna, Austria}

\date{\today}

\begin{abstract}
Molecular Dynamics simulations are reported for the structural and
thermodynamic properties
of submonolayer xenon adsorbed on the $(111)$ surface of platinum for
temperatures up to the (apparently incipient) triple point and beyond.
While the motion of the atoms in the surface plane is treated with a
standard two-dimensional molecular dynamics simulation, the model takes into
consideration the thermal excitation of quantum states associated with
surface-normal dynamics in an attempt to describe the apparent smoothing of
the corrugation with increasing temperature.  We examine the importance of
this thermal smoothing to the relative stability of several observed and
proposed low-temperature structures.  Structure factor
calculations are compared to experimental results in an attempt to determine
the low temperature structure of this system.  These calculations provide
strong evidence that, at very low temperatures, the domain wall structure of
a xenon monolayer adsorbed on a Pt$(111)$ substrate possesses a chaotic-like
nature, exhibiting long-lived meta-stable states with pinned domain walls,
these walls having narrow widths and irregular shapes.  This result is
contrary to the standard wisdom regarding this system, namely that the very
low temperature phase of this system is a striped incommensurate phase.
We present the case for further experimental investigation of this
and similar systems as possible examples of chaotic low temperature phases
in two dimensions.
\end{abstract}



\maketitle

\section{\label{intro}Introduction}

The physical adsorption of gas atoms (adatoms) on solid crystalline
substrates produces an
interesting collection of phase transitions in a nearly two-dimensional
(2D) environment.\cite{BruColZar97} These include a rich variety
of structural phase transitions driven by mismatches between the natural
periodicities of the adsorbed layer and the crystalline
substrate.\cite{BruDieVen07,CorBonPol08}
These effects are not restricted to the lowest layer, but can be found even
in upper layers as they are subjected to the periodic field generated by the
lower ones.\cite{CorBonPol08,AhnLeeKwo16,Mor17}

Of the various possible combinations of adsorbates and adsorbents, the
adsorption of noble gas atoms on well-ordered crystalline surfaces provides an
especially attractive and useful set of examples for both theorists and
experimentalists interested in fundamental questions.\cite{MadVit17}
There are a number of reasons to consider these excellent examples of
model systems.  First,
the interactions between the various noble gas atoms in vacuum are
simple and well understood\cite{Azi11} with the modifications to these
interactions due to the adsorption of the atoms on certain surfaces being
reasonably well modeled.\cite{Bru83} Second, there is a well-developed
approach to the interaction of these atoms with the crystalline
surfaces\cite{Ste74} with the values of the interaction parameters
(for many specific systems) being reasonably well
determined.\cite{VidIhmKim91} Therefore, there is the opportunity to make
quantitative comparisons between theory and experiment with some confidence,
more than may be justified for many other adsorbate-adsorbent combinations
(where the modeling is not on as firm  a foundation and the systems are
not as well characterized).\cite{JarDwoFou04,HayMan12,FarRie98}

The adsorption of a xenon adatom on the $(111)$ surface of platinum
(Xe/Pt) is one of the more interesting cases of physical
adsorption.\cite{GelBakHol16,CheAlSJoh12,DieSeyCar04}
Unlike many of the cases involving the physical adsorption of noble gas
atoms on crystalline substrates, the sites for adsorption are directly
``over'' the Pt surface atoms and not at the hollow
positions,\cite{DieSeyCar04,SeyCarDie99}
the latter being the expected situation for dielectric surfaces and confirmed
for Xe on the basal plane surface of graphite (Xe/Gr).
Perhaps related to this, the corrugation along the surface plane is much
larger than it is for many other cases of the adsorption of noble
gases.\cite{BarRet92,*BarRet94}
This strong bonding of the xenon adatom to the platinum surface is
associated with a significant redistribution of the electron density in
both the adatom and the platinum surface atoms, which gives this system
some characteristics of chemisorption.\cite{BetBir00}

The strong corrugation and dilated lattice of Xe/Pt suggests an
interesting question: Is there anything unusual about the structure of
this system at very low temperatures?
In chemisorbed systems, there exist examples exhibiting chaotic structures
at very low temperatures.\cite{LiaDAmLee90} Paradoxically, at higher
temperatures, such systems relax to form a more regular, ordered structure.
Similar behavior has been seen in some magnetic systems\cite{JenBak83} and
in systems with charge-density waves.\cite{Bak82}
In all these cases, the resulting chaotic structure can be attributed to
pinned domain walls which lock down the structure with a certain degree of
randomness in the placement of these walls.\cite{Bak82}
In adsorbed systems, this chaotic behavior relies on the forces of the
substrate lattice acting on the domain walls (the Peierls pinning
force)\cite{Hob65a,Bak82,Hob65,Zim72,Joo90,BouBru04,JenBak83}
being comparable to or stronger than the interactions between these domain
walls.\cite{Bak82}  This same physics appears to be important to a
related problem, the development of friction in monolayer patches sliding
across periodic substrates.\cite{VarVanGue15,ManDavVan17,GueWijMer17}

Here we investigate whether similar chaotic behavior at low-temperatures
can be present in the physical adsorption case of Xe/Pt.
In order to do so, we perform molecular dynamics simulations using a
hybrid approach, combining a classical treatment of the dynamics
along the surface with a quantum treatment of the dynamics perpendicular to
that surface. We refer to this as a quasi-two-dimensional (Q2D) treatment,
justifying this treatment based upon the very different nature of the
variation in the surface interaction along the surface compared to
the variation perpendicular to that surface.
Our model is constructed using a combination of the Barker-Rettner
model\cite{BarRet92,*BarRet94} for Xe/Pt with a Xe-Xe
``Hartree-Fock-Dispersion'' interaction,\cite{Azi11}
modified by the McLachlan interaction for Xe/Pt.\cite{Mcl64}
The largest case study reported here approaches the size of some experimental
systems, but is smaller than the experimental best-case
scenario.\cite{KerDavPal86,Ker94}

We provide evidence of irregular, extremely narrow domain walls for the
low temperature Xe/Pt system.  These walls tend to zigzag in a rather
haphazard (and perhaps ``chaotic'') fashion and do not appear
to relax as the run is extended in time, nor when the temperature is raised.
This behavior is consistent with these domain walls being pinned, at low
temperature, by the Peierls force, and is in agreement with early preliminary
calculations.\cite{Note-Stud}  We will label such structures as chaotic,
although we can not show they fit any strict definition of such a state.
We will use the phrase ``disordered state'' to refer to the phase above the
melting transition.\cite{NovBruBav15}

We also report other structural and thermodynamic analyzes for constrained
(the xenon monolayer uniformly fills the entire simulation cell) and
unconstrained (a xenon patch in center of the simulation cell) system
geometries. This includes calculations of substrate
corrugation parameters, determination of the ground state phase, and evidence
of meta-stability for the low temperature phases. The effects of size
dependence are made explicit by the calculation of both $\psi_0$
(the hexatic order parameter)\cite{NovBru14} and $\psi_6$
(the Net-Domain-Phase order parameter)\cite{NovBruBav15} as functions of the
temperature and size of the system. The effects of the rotation of the
monolayer with respect to the substrate are also examined. Some of these
results are to be found in the supplementary material.       % \cite{NovBavSM17}

Our calculations of the static structure factor allow for comparisons
between our simulations and known experimental results.
There is a claim, based upon HAS experiments,\cite{KerDavZep88} that the very
low temperature phase of Xe/Pt is an striped incommensurate phase. This
has become the conventional wisdom for this
system.\cite{BruDieVen07,ComKerPoe92}
The principle experimental evidence for this striped phase
involves an analysis of the static structure factor,
comparing the hexagonal domain wall structure to that of the
striped phase.\cite{KerDavPal86,KerDavPal86f,KerDavZep87,Ker87}
However, it does appear that this analysis did not consider
the possibility of a chaotic (i.e.\ disorganized) domain structure
that is made evident in our simulations,
and as such, makes no prediction about the existence of such a state.

We show that chaotic-like structures can
exist as meta-stable (long-lived) states in Xe/Pt.
Furthermore, the resulting structures can
mimic the experimental results used as evidence for the striped phase.
In addition, some aspects of the experimental results seem at odds with
the structure factor for the meta-stable striped phase reported here.
Some results of our simulations have been reported in
Ref.~[\onlinecite{NovBruBav15}];
this article being both a follow-up to and a completion of that work.

\section{\label{Model}Model for X\lowercase{e}
 Adsorbed on P\lowercase{t}$(111)$}

Much of the behavior seen in the simulations of this system is driven by
the strong corrugation and the dilated lattice of Xe/Pt.  The minimum barrier
to translation from one adsorption site to the next is roughly $275$~K,
whereas the minimum in the effective interaction between xenon atoms is about
$238$~K.\cite{BruNov00}  In contrast, the minimum barrier to
translation for Xe/Gr is about $50$~K, while the Xe-Xe interaction is
nearly unchanged.\cite{BruNov08,NovBru14} In addition, the Xe-Xe spacing for
the $\sqrt{3} \times \sqrt{3}$\,R\,$30^{\circ}$ (\Rt{3}) phase of Xe/Pt is
4.80~\AA, significantly larger than the position of the minimum in
the Xe-Xe interaction (4.37~\AA).  It would seem that this particular
combination of a large corrugation and a dilated \Rt{3} lattice is what leads
to a replacement of a normal triple-point transition with an order-disorder
transition (an incipient triple-point).\cite{NovBruBav15}
We did not explore the boundaries of the parameter space that would generate
this behavior.

The literature for classical simulations of the Xe/Pt monolayer, both
Molecular Dynamics (MD) and Monte Carlo (MC), is quite sparse.  There
are some early MD simulations of small
systems,\cite{BlaBop87,BlaBop86,BlaJan89,BlaJan89a,BruSchFed02}
but no MC work to speak of.
There is an extensive body of work (by Bruch and Gottlieb) on the stability of
various possible structures that might exist in this sort of system (using a
harmonic lattice dynamics approach). In particular, there is a direct
application of their ideas to the Xe/Pt system,\cite{Got90,GotBru91b,GotBru91e}
Unfortunately, some of these early calculations used older (and less realistic)
forms for the Xe-Xe interactions and-or simplistic models for the Xe-Pt
interactions.  However, the stability of the various possible structures
for this system has been shown to be sensitive to relatively small changes
in these interactions.\cite{NovBruBav15,Got90,BruSchFed02}

A model that has been successful in describing the interaction of a xenon
atom with the platinum $(111)$ surface is that of Barker and Rettner
(BR),\cite{BarRet92,*BarRet94} a semi-empirical model that fits a significant
collection of experimental data.  Only a few of the early calculations for
Xe/Pt used
the BR model for this interaction, and more importantly, these calculations did
not directly examine the effects of thermal excitation perpendicular to the
surface on the effective corrugation of the system.  This motion has an
important effect on the thermal smoothing of the corrugation and thus on the
thermal properties of the monolayer.\cite{NovBruBav15} Furthermore, when the
dynamics of this (and other) adsorbed systems in the surface-normal direction
has been treated, it has typically been done by using a purely classical
treatment of that motion.\cite{FleEtt02,FleEtt06,BruSchFed02,UstDo14,Ust14}

The problem with a purely classical approach to the surface-normal dynamics
is that even when the surface-parallel motion is well treated by classical
dynamics, the same can not be said of the surface-normal dynamics.  This is due
to the narrowness of the potential energy well in the surface-normal direction
and the corresponding large excitation energies of the adatom.  The
corresponding thermal motion has an important influence on the effective
corrugation and thus on the predictions of stable structures and phase
transitions in this system.\cite{BouBru04,BruSchFed02,NovBruBav15}
In the following, we describe how our model is constructed to
overcome these issues.

\subsection{\label{Model:Q2D}Q2D approach}

The adatom coordinates along the surface plane are $(x,y)$
(denoted by $\bm{r}$), while the coordinate in the surface-normal direction
is $z$.  We start with a quantum description of the system, approximating the
exact wave function by a set of product wave functions having the
form:\cite{Nov92}
\begin{equation}
\Psi(\bm{r}_1,z_1,\bm{r}_2,z_2, ... ) =
                \Psi_{\parallel}(\bm{r}_1,\bm{r}_2, ... ) \times
                          \Psi_{\perp}(z_1,z_2, ... ) .
\end{equation}
More to the point, we consider the Hilbert space of all such
functions, assuming appropriate orthogonality and completeness conditions for
the set.  It must be noted that, in the end, we will approximate the dynamics
of the surface parallel terms using a classical MD simulation, but will
retain a quantum description of the surface-normal dynamics as noted in
the previous paragraph.

Following Ref.~[\onlinecite{Nov92}], it is convenient
to consider three subsets of contributions to the total energy of the system.
These three energy contributions
are denoted by $E_z$, $E_{xy}$, and $E_{xyz}$ (each on a per adatom basis).
The first contribution, $E_z$, is the kinetic energy associated with the
$z$-direction plus the laterally averaged substrate interaction $U_0(z)$.
This term depends only on the $\Psi_{\perp}$ factor, and it is the
thermal behavior of this factor that is primarily responsible for the
temperature dependence of the substrate corrugation.  The second contribution,
$E_{xy}$, is the remaining kinetic energy terms plus the interaction between
the xenon adatoms. Strictly speaking, this interaction term depends
upon both the $\bm{r}$ and $z$ variation of the wave function.  However,
it is a very good approximation to treat this as dependent only on the
$\Psi_{\parallel}$ factor because of the narrowness of the $\Psi_{\perp}$
functions.\cite{Nov92}  The third and final term consists of the remaining
contributions to the Xe/Pt interaction, that is the non-zero $\bm{G}$
(platinum reciprocal lattice vectors) terms in a Fourier expansion of
the BR interaction, projected onto the $\bm{r}$ plane as described
below and in the supplementary material.                    %  \cite{NovBavSM17}
This term, which determines the effective corrugation,
depends on both the $\Psi_{\parallel}$ and the $\Psi_{\perp}$ factors in
$\Psi$.  In this work, as in Ref.~[\onlinecite{Nov92}], $\Psi_{\perp}$
is written as a product of single-particle Gaussians, effectively treating the
monolayer vibrational mode polarized in the surface-normal direction as a
flat mode with no variation in frequency across the two-dimensional Brillouin
zone.\cite{BruNov00,Nov92} This has been shown to be a good
approximation.\cite{BruNov00}

 This Q2D approach involves an explicit assumption that the adatom
finds the optimum $z$-position as it moves along the surface. This
means that there is an implicit assumption being made about the coupling of
the motion in $z$ to that in $\bm{r}$.
Given this, there are a number of avenues to the projection of the 3D potential
energy of the BR model into the plane of $\bm{r}$, some purely classical
in approach and some quantum in nature.  The quantum projections build
upon the classical projections by averaging various expressions of the
classical projections over the zero-point (and thermal) motion of the
adatom in the $z$-direction.
That is, a quantum projection corresponding to any particular
classical one replaces the potential energies (and their derivatives with
respect to the $z$-displacement) with the appropriate quantum thermal
averaging using the Self-Consistent Phonon (SCP) Gaussian distributions as
specified in Refs.~[\onlinecite{Nov88,Nov92}].
This Q2D approach results in a modified form of the 3D Steele expansion
of the potential energy of an atom due to the surface of a crystalline
substrate,\cite{Ste73,Ste74} using quantum thermal averaging to project the
3D potential energy into the plane of $\bm{r}$.
This effective potential energy, denoted by $\tilde{U}(\bm{r})$, can be
written as a Fourier series in the form:
\begin{equation}
\label{SteeleEqQ2D}
	\tilde{U} (\bm{r}) =
	\sum_{ \bm{G} } \tilde{U}_{\bm{G}} \exp ({ i \bm{G} \cdot  \bm{r}} ),
\end{equation}
where $\bm{G}$ is a reciprocal lattice vector of the two-dimensional
surface lattice and the effective, Q2D Fourier coefficients
$\tilde{U}_{\bm{G}}$ depend upon temperature as a result of the quantum
thermal averaging of the xenon dynamics in the surface-normal direction.
(See Appendix~\ref{IndexMapping} for the reciprocal lattice naming and
indexing conventions used).

Our Q2D approach uses the quantum states that describe the $\Psi_{\perp}$
factor to calculate the Q2D Fourier coefficients that
describe the variation of the substrate corrugation as a function of $\bm{r}$.
The details of how this is done is
described in the supplementary material,                     % \cite{NovBavSM17}
which describes two quantum-based methods and three classical approximations.

There is an important caveat in this approach, and it is associated
with the mixing of a classical treatment of the $xy$-motion with a quantum
treatment of the $z$-motion.  It is obvious how to deal with both the $E_z$
and the $E_{xy}$ terms since the first depends only on $z$ (quantum treatment)
and the second only on $\bm{r}$ (classical treatment).  However, the $E_{xyz}$
term depends upon both, so there is some ambiguity about how to properly treat
this term because the effective Fourier coefficients defined by this term can
reasonably be averaged over both the $\bm{r}$ and the $z$ motions.\cite{Nov92}
The decision was to match the MD and SCP energies (at zero temperature) as
closely as possible by following the procedure in Ref.~[\onlinecite{Nov92}],
even though this might overstate the effects of quantum and thermal smoothing
at finite temperatures. On the other hand, this approach does come close to
aligning the adatom-substrate classical potential energy with the
corresponding SCP potential energy, even at finite temperatures.  This approach
can be interpreted as an approximate wave packet calculation, the
approximations involving the replacement of the quantum thermal average of
the Fourier term with a cumulant expansion as is done in
Ref.~[\onlinecite{Nov79}], and the use of constants for the second cumulants
of the Gaussian distributions, these cumulants being calculated separately
by a SCP treatment of the \Rt{3} phase.\cite{BruNov00}             % NovBavSM17}
Additional details are to be found in the supplementary material.

\subsection{Xenon-xenon interaction}

The interaction between two isolated xenon atoms is taken to be the
HFD-B2 interaction on page 177 of Ref.~[\onlinecite{DhaMeaAll90}],
which is a ``Hartree-Fock-Dispersion'' interaction.\cite{Azi11}
This interaction does an excellent job of describing the various features
of the xenon-xenon pair interaction in vacuum.
However, since the Xe atoms are adsorbed on a surface, there is
a modification of this pair interaction generated by the dielectric
properties of this surface.
Thus, the HFD-B2 interaction is modified by adding a McLachlan
interaction\cite{Mcl64} with the parameters given by Bruch for the
Pt$(111)$ surface.\cite{BruGraToe00}
This model (HFD-B2+McLachlan) has been used successfully
in Ref.~[\onlinecite{BruNov00}] for a lattice dynamics analysis of
this system and in Ref.~[\onlinecite{BruGraToe00}] as part of the analysis
of HAS experiments.  In addition, as was the case in
Ref.~[\onlinecite{BruNov00}], the effects of any
induced xenon dipole as well as three-body terms are ignored. Details,
justification, and values of relevant parameters are given
in Ref.~[\onlinecite{BruNov00}] as is the justification for ignoring any
dipole-dipole interactions.  The parameter values in
Ref.~[\onlinecite{BruNov00}]
are based on the work in Ref.~[\onlinecite{BruGraToe00}].

\subsection{Xenon-platinum interaction}

The position of the preferred adsorption site, which is over the surface
Pt atom, has an important effect on the stability of the various phases
of Xe/Pt.\cite{RejAnd93,BarRet92,*BarRet94,WeaStiMad97,Got90,GotBru91e,
BruDieVen07,SeyCarDie98}
We choose the BR model to capture this aspect of the Xe/Pt system,
as well as other
important physical characteristics.\cite{BarRet92} Furthermore, it is
considered to be one of the more successful interaction models for this
system, with much to recommend it.\cite{BruColZar97,BruDieVen07}
It has been used in a successful treatment of the lattice vibrations for
this system,\cite{BruNov00,BruGraToe00,BruGraToe98} and
the region of the potential well that is most important to lattice
dynamics calculations has a similar importance here.
The details of the implementation of this interaction model,
including the parameters used for these calculations,
are identical to those used in the lattice dynamics calculations
of Ref.~[\onlinecite{BruNov00}].

\subsection{\label{Model:MD}Molecular dynamics simulations}

The simulations presented here are standard molecular dynamics
simulations in 2D (fixed particle number, area, and total energy)
with the substrate potential energy given by Eq.~(\ref{SteeleEqQ2D}).
The simulations are carried out using scaled equations with a length scale
of 4.3656~\AA, an energy scale of 282.8~K, and a time scale of 3.262~ps.
The technical details of these simulations are found in the supplementary
material.                                                     %\cite{NovBavSM17}
We have carried out a series of simulations for
different sets of these Fourier coefficients,  using a range of coefficients
which should bracket the most likely values both at low temperatures and
at high temperatures.  It is our expectation
that we have a bracket around the most likely behavior of the system
for the range of temperatures of interest. Details and supporting
arguments are found in Sec.~\ref{Results:Q2D} and in the supplementary
material.                                                     %\cite{NovBavSM17}

\begin{table}
\caption{\label{Table:CaseStudies} Parameters that define typical
case studies discussed in this work.  All energy values are in kelvin.}
\begin{ruledtabular}
\begin{tabular}{ l | c | c | c | d }
Case Study & Substrate & Projection & Size & U_{(10)} \\
\hline
BR:65K\footnotemark[1] & Pt$(111)$ & BR & 65K &  -35.6 \\
BR-H:20K\footnotemark[2] & Pt$(111)$ & BR-H & 20K &  -35.6 \\
U25-H:20K\footnotemark[2] & Pt$(111)$ & U25-H & 20K &  -25.0 \\
\end{tabular}
\end{ruledtabular}
\footnotetext[1]{Constrained geometry with 65536 adatoms.}
\footnotetext[2]{Unconstrained geometry with 20064 adatoms.}
\end{table}

We refer to the selections of different projections and system sizes as
separate case studies, each being tagged using a notation that consists of
two strings separated by a colon.  The first string specifies the corrugation
model for Xe/Pt and the second specifies the size of the system.  An example
is the case study U25:65K, where the U25 refers to a Fourier expansion with
a single independent amplitude ($U_{(10)} = -25$~K) and the 65K refers to
the system having $65536$ particles in the box.  A BR for the first string
denotes the Barker-Rettner interaction as in cases 1 and 2 of
Table~\ref{Table:CaseStudies}.  Furthermore, if the first string is
terminated by a ``-H'', as in BR-H, then that infers the case study is
for a unconstrained geometry (hexagonal shaped patch). A ``-U'', as in BR-U,
denotes an initial uniaxial configuration (striped phase) which was
either unconstrained (i.e.\ rectangular patch for the 20K size) or a
constrained geometry (i.e.\ for the 65K size).  If there is no such
designation, then the simulation is for a constrained geometry with the
initial lattice being either hexagonal or centered-rectangular.  The number
density of the \Rt{3} structure, denoted by $\rho_0 = 0.05016$~\AA$^{-2}$,
is used to scale the density in various figures and tables.

\section{\label{Results}Results}

The results of our simulations are organized as follows.
In Sec.~\ref{Results:Q2D}, we report and discuss
the calculation of the parameters of the substrate corrugation.
In Sec.~\ref{Results:MD}, we define the different initialization phases
that were studied, followed by an analysis of the ground state structure 
for constrained geometries in Sec.~\ref{Results:MD-I} and for unconstrained 
geometries in Sec.~\ref{Results:MD-II}.  In Sec.~\ref{Results:MD-Structure},
we report calculations and analysis of the structure factor of the different 
case studies, which is the figure of merit for the comparison to experimental 
results, reported in Sec.~\ref{Comparison}.  Additional explanations, results,
comparisons, and discussion is to be found in the supplementary
material;                                                    % \cite{NovBavSM17}
such as the details of the calculation of the
corrugation that is the basis for the conclusions of
Ref.~[\onlinecite{NovBruBav15}].

\subsection{\label{Results:Q2D}Substrate corrugation}

\begin{table}
\caption{\label{Table:CaseStudyList} The Fourier coefficients used in these
MD calculations.
All values are in kelvin.}
\begin{ruledtabular}
\begin{tabular}{ l | d | d | d }
Projection & U_{(10)} & U_{(11)} & U_{(20)} \\
\hline
BR\footnotemark[1]  & -35.64 & 0.39 & 0.48 \\
UN\footnotemark[2] & -N.00 & - & - \\
\end{tabular}
\end{ruledtabular}
\footnotetext[1]{Calculated using the classical perturbation approach.}
\footnotetext[2]{N $\in$ \{20, 25, 30, 35\}.}
\end{table}

The descriptions and the corresponding Fourier coefficients used for the case
studies examined here and in Ref.~[\onlinecite{NovBruBav15}] are given in
Tables~\ref{Table:CaseStudies} and \ref{Table:CaseStudyList}.
The naming and indexing conventions used for the reciprocal lattices
are described in Appendix~\ref{IndexMapping}.  The details of these
calculations are to be found in the supplementary material.  % \cite{NovBavSM17}
The origin (or zero) of the 2D energy calculation for a given case study is
the corresponding value of $E_z$. These values are given in
Table~\ref{Table:Projections} for three temperatures, along with the
minimum value of the laterally
averaged substrate potential energy $U_0$ and the corresponding values of
$z_{opt}$, the optimum value of $z$.
\begin{table}
\caption{\label{Table:Projections} Parameters for the projection of the
	BR model onto the surface plane. The minimum in the laterally
	averaged potential
        well is $U_0$, the 1D total energy is $E_z$ (includes SCP
	zero-point and thermal energies), and the corresponding optimum
	$z$-position for the laterally averaged substrate potential energy is
	$z_{opt}$.  All energies are in units of $10^3$ kelvin and all
	distances are in \AA.}
\begin{ruledtabular}
\begin{tabular}{ l | d | d | d }
Approach & U_0 & E_z & z_{opt}\footnotemark[1] \\
\hline
Classical & -2.746 & -2.746 & 3.43  \\
Quantum\footnotemark[2]& -2.738 & -2.730 & 3.44 \\
Quantum\footnotemark[3]& -2.715 & -2.684 & 3.48 \\
Quantum\footnotemark[4]& -2.691 & -2.636 &  3.52 \\
\end{tabular}
\footnotetext[1]{Optimized using only the $U_0$ term.}
\footnotetext[2]{$T = 0$.}
\footnotetext[3]{$T = 60$.}
\footnotetext[4]{$T = 110$~K.}
\end{ruledtabular}
\end{table}
As described in Sec.~\ref{Model:Q2D}, the quantum optimization uses SCP
averaging, thus minimizing the free energy contribution associated with
the $z$-motion.
\begin{table}
\squeezetable
\caption{\label{Table:BindingEnergy} Theoretical and experimental values
for the binding energy $\epsilon_0$, the isosteric heat of adsorption
$q_{st}$, and the equilibrium separation of the
isolated xenon atom from the platinum surface $z_{opt}$.  The theoretical
values are obtained by using up to three independent $U_{\bm{G}}$ coefficients
with the SCP approximation for the $z$-wise single particle dynamics.
All energies are in units of $10^3$ kelvin and all distances are in \AA.}
\begin{ruledtabular}
\begin{tabular}{ l | d | d | d }
Source & \epsilon_{0} & q_{st} & z_{opt} \\
\hline
Theory\footnotemark[1] & 2.67 & 2.79 & 3.50 \\
% Theory\footnotemark[2] & 2.72 & 3.10 & 3.38 \\
Theory\footnotemark[2] & 2.70 & 2.90 & 3.41 \\
Kern\footnotemark[3]   & - & 3.21 - 3.31 & - \\
Diehl\footnotemark[4]   & - & 3.02 - 3.25 & 3.40 \\
\end{tabular}
\end{ruledtabular}
\footnotetext[1]{This work: $q_{st}$ for a 2D ideal gas using
	$U_0$ and $T = 80$~K.}
\footnotetext[2]{This work: $q_{st}$ for a 2D lattice gas of
	3D-Osc. with $T = 70$~K.}
\footnotetext[3]{Ref.~[\onlinecite{KerDavZep88}]:
	$T \ge 70$~K and coverage $\Theta \le 0.03$.}
\footnotetext[4]{Ref.~[\onlinecite{DieSeyCar04}]: $T = 110$~K.}
\end{table}

For comparisons with experiment, we show the binding energy $\epsilon_0$,
the isosteric heat $q_{st}$, and the optimum $z$-position $z_{opt}$ for the
BR model as well as corresponding  experimental values.
These results are found in Table~\ref{Table:BindingEnergy}.
The relations between $q_{st}$ and the theoretical
$\epsilon_0$ are explained in the supplementary material.    % \cite{NovBavSM17}

The theoretical and experimental values of $z_{opt}$ agree very well, while
the corresponding values of $q_{st}$ differ by about 5~to~15~\%.  The
calculated values of the $z$-wise vibrational amplitude of the xenon motion
for the \Rt{3} phase at 110~K are in excellent agreement with experiment,
theory giving {0.16~\AA}\cite{BruNov00,Note-Novaco} and experiment
giving 0.17~\AA.\cite{DieSeyCar04} In all, the agreement between the BR model
results and the experimental results is both respectable and satisfactory.

\begin{table}
\caption{\label{Table:Q2D-CP} The Fourier coefficients calculated using
	the classical projections described in the supplementary 
	material.                                            % \cite{NovBavSM17}
	All values are in kelvin.}
\begin{ruledtabular}
\begin{tabular}{ l | d | d | d }
Projection & U_{(10)} & U_{(11)} & U_{(20)} \\
\hline
Method 1 & -34.16 & -  & - \\
Method 2 & -33.37 & -0.80 & -0.54 \\
Method 3 & -35.64 &  0.39 &  0.48 \\
\end{tabular}
\end{ruledtabular}
\end{table}
A comparison of the results for the calculation of the finite $\bm{G}$
Fourier coefficients, using the classical approach from
Sec.~\ref{Model:Q2D} and the supplementary material,         % \cite{NovBavSM17}
is displayed in
Table~\ref{Table:Q2D-CP}.  The first method produces a good estimate of
$U_{(10)}$, quite compatible with that of the second method.  The second
method produces values of $U_{(11)}$ and $U_{(20)}$ that are rather small,
but perhaps not completely negligible. The third method has a slightly
different set of $U_{\bm{G}}$, but an overall corrugation that is not that
different than the other two.

\begin{table}
\caption{\label{Table:Q2D-QP} The Fourier coefficients calculated using
	the quantum projections described in the supplementary
	material.                                            % \cite{NovBavSM17}
	The calculations were carried out for temperatures of $0$,
	$60$, and $110$~K. All values are in kelvin.}
\begin{ruledtabular}
\begin{tabular}{ l | d | d | d }
Projection & U_{(10)} & U_{(11)} & U_{(20)} \\
\hline
Method 1\footnotemark[1] & -32.7 & - & - \\
Method 2\footnotemark[1] & -31.9 & -0.8 & -0.6 \\
\hline
Method 1\footnotemark[2] & -29.01 & - & - \\
Method 2\footnotemark[2] & -27.49 & -0.94 & -0.87 \\
\hline
Method 1\footnotemark[3] & -26.08 & - & - \\
Method 2\footnotemark[3] & -23.85 & -1.18 & -0.83 \\
\end{tabular}
\end{ruledtabular}
\footnotetext[1]{Calculated at 0~K.}
\footnotetext[2]{Calculated at 60~K.}
\footnotetext[3]{Calculated at 110~K.}
\end{table}
Comparisons using the corresponding quantum projections      % \cite{NovBavSM17}
are displayed in Table~\ref{Table:Q2D-QP}.  At zero kelvin, there is a
small reduction in the magnitude of the
classical value of the Fourier coefficient for $\tilde{U}_{(10)}$ due
to quantum effects. There are corresponding small changes in the (absolute)
values of the others.  At $110$~K, the reduction in $U_{(10)}$ is
significantly larger but still with small values for the others.
Given these results, it would be reasonable to use a range of
$-35.0$~K to $-24.0$~K as the appropriate one for $U_{(10)}$ values in the
temperature range from zero to a bit over $110$~K.  However, this assumes that
the adatom is able to maintain the optimal distance from the surface, even at
the highest temperatures; and also assumes that there are no effects generated
by the thermal excitation of the Pt surface itself
(see the supplementary material).                            % \cite{NovBavSM17}
Without a direct calculation of these
effects, we can only guess how important these might be. However, such a
calculation would go far beyond the goals of this work.  Instead, we have
simply used an arbitrary lowering of the corrugation, using a $U_{(10)}$ of
$-20.0$~K as a corrugation lower bound.\cite{NovBruBav15}
This approaches the smallest corrugation that is reasonable, based
upon the results above and an observed \Rt{3} phase that
is stable in the temperature range from $60$~K through over $110$~K.
Lowering the corrugation too much destabilizes the \Rt{3} phase over much of
that temperature range.  Our lower bound preserves this stability, but it does
represent a significant lowering of the corrugation. In a couple of special
test cases, a $U_{(10)}$
of $-15.0$~K was also used, but nothing differed in any interesting way.
These calculations and the values found in Tables~\ref{Table:Q2D-CP} and
\ref{Table:Q2D-QP} were the ones used in Ref.~[\onlinecite{NovBruBav15}].

Finally, calculations were carried out comparing theoretical and
experimental root-mean-square (RMS) vibrational amplitudes of the xenon
atoms in an attempt to better constrain the corrugation. However, the results
were not useful in improving our estimate of this corrugation. These
calculations and the corresponding results are described in the supplementary
material.                                                    % \cite{NovBavSM17}
Some additional consequences of this
smoothing of the corrugation with increasing temperature are to be found in
Ref.~[\onlinecite{NovBruBav15}].  As shown in that paper, reductions in
the corrugation are important to the understanding of the phase transitions in
the Xe/Pt system.

\subsection{\label{Results:MD}Molecular dynamics}

The implementation of the basic MD simulation is outlined in the supplementary
material,                                                    % \cite{NovBavSM17}
and closely follows
Refs.~[\onlinecite{BruNov08,NovBru14}].  Using these simulations along
with the free
energy analysis as discussed in the supplementary material,  % \cite{NovBavSM17}
we examine the stability, structure, and thermal behavior of the \Rt{3} phase,
the striped incommensurate (SIC) phase, and the hexagonal incommensurate (HIC)
initializations. Both aligned (AIC) and rotated (RIC) HIC structures
are examined.  In addition, the stability of structures having irregular
and apparently pinned domain walls are examined and compared to the others.
These ``chaotic'' structures will be referred to as ``chaotic hexagonal'' and
``chaotic striped'', even though it is not clear if they satisfy the strict
definitions of a chaotic state.  We examine these structures by initializing
the system in these configurations and then following the simulations out to
apparent thermal equilibrium.  When possible, free energy comparisons are
made between the various structures to determine the stable phase (this can
not be done for the constrained geometries).  Some initializations into these
structures produce stable (or meta-stable) disordered (chaotic) domain-wall
structures.
Comparisons of the details of this work to that of
Refs.~[\onlinecite{BruNov08,NovBru14}] are described
in the supplementary material.                               % \cite{NovBavSM17}
In addition to the
thermodynamic functions calculated in Ref.~[\onlinecite{NovBru14}],
calculations were carried out for
the specific heat at constant area, and the two order parameters:
$\psi_6$ and $\psi_0$.
The definitions and thermal behavior of these two order
parameters are to be found in Refs.~[\onlinecite{NovBru14,NovBruBav15}].
Other details of
the thermodynamic calculations as well as corresponding results are to
be found in that same reference.  Further details follow here and in
the supplementary material.                                  % \cite{NovBavSM17}

\subsubsection{\label{Results:MD-I}Constrained geometry}

Our MD simulations indicate that the ground state of the BR model is the
\Rt{3} state.  Furthermore, at finite misfits, the HIC phase is the stable
phase. However, for small misfits, the SIC phase will strongly compete with
the HIC phase and have nearly the same free energy (see the supplementary
material                                    %  in Ref.~[\onlinecite{NovBavSM17}]
for details).  If the low temperature
corrugation is smoother than about $U_{(10)} \approx -30$~K (the
exact value dependent upon the importance of quantum effects), the IC
phase becomes the stable low temperature phase and the ground state of
the system.  It is not unreasonable to speculate that, depending
upon the actual smoothing due to quantum and thermal effects, (and
corrections to the BR model) it might be possible
for this system to have a \Rt{3} ground state with significant competition
from a SIC phase, a stable IC phase at higher temperatures (but less than
about 60~K), and then a return to a stable \Rt{3} phase at roughly that
temperature.

If the actual corrugation is well below the values above, but still
stronger than about $U_{(10)} \approx -15$~K, the system would evolve
from an IC low temperature phase to the \Rt{3} phase as the temperature
increases. However, the temperature range of stability for the \Rt{3} phase
is reduced as the corrugation decreases.  For corrugations much lower than
this $U_{(10)} \approx -15$~K lower limit, the system does not enter the
\Rt{3} phase before the disordering temperature, but rather remains in the
IC phase until melting.  Preliminary calculations of quantum corrections do
not appear to alter this conclusion in any significant manner (the
calculations indicating that the AIC and SIC states have nearly identical
free energies at small misfits).\cite{Note-Novaco}

\subsubsection{\label{Results:MD-II}Unconstrained geometry}

Most simulations for the unconstrained geometry were carried out for a
system size of $20$K.  The thermodynamic behavior of the unconstrained
geometry, in the (average) density range of $0.14$ to $0.67$ times
$\rho_0$, is not very sensitive to variations in that density (provided
the system is an isolated, single patch).  Most of the data was taken
with an average density of roughly $0.45$ to $0.55$ on this scale, but
some simulations at the highest and lowest densities were used so
that the sensitivity of the results to changes in the average density
could be examined. Nothing of significance was found.

The thermodynamic stability analyses (see supplementary material) of the
various meta-stable structures in this system                % \cite{NovBavSM17}
demonstrates that the \Rt{3}
phase is the stable phase of both the BR-H and U30-H projections at
low temperatures, but that the SIC phase has nearly the same energy
at low temperatures.  Thus, while the \Rt{3} phase is the expected low
temperature phase of the BR model for Xe/Pt, it will have strong competition
from the SIC phase. This may be what drives the system into the observed
``chaotic'' domain structure that is observed in the simulations.
Furthermore, even though simulations using the smaller corrugations in
Table~\ref{Table:CaseStudyList} show that the stable phase at very low
temperature would be an IC phase, the stable phase at higher temperatures,
even for these smaller corrugations, is still the \Rt{3} phase.  Using the
corrugation values shown in Table~\ref{Table:CaseStudyList}, the stable phase
at temperatures just below the melting temperature does appear to be
the \Rt{3} phase.
The effects of quantum behavior and the implications for the
determination of the stable state will be addressed in a future
publication.\cite{Note-Novaco} However, preliminary calculations
indicate that the basic conclusions of the MD stability analysis shown here
are not significantly altered by quantum effects as calculated by a SCP
type analysis.\cite{Nov92,Note-Novaco} Finally, the transition
from \Rt{3} ground state to IC ground state occurs at
values of $U_{(10)}$ between $-30$ and $-25$ kelvin, the actual value
most likely closer to $-25$ than $-30$ kelvin. However, this value is affected
by quantum corrections and it needs a more careful examination. This will also
be addressed in a future publication.\cite{Note-Novaco}

\subsubsection{\label{Results:MD-Structure}Structure of the monolayer}

The complexity of the monolayer structure makes its description
somewhat difficult and cumbersome. Nevertheless, there are a number of
characteristics of the monolayer that are useful to describe in detail,
and this can be done with some confidence.  Some of these details
are to be found in the supplementary material,               % \cite{NovBavSM17}
others follow.

The structure factor for the \Rt{3} phase shows the expected behavior for both
constrained and unconstrained geometries.  At low temperatures, the widths of
the $S(\bm{Q})$ peaks are consistent with the size of the system, and show the
expected decrease in peak height and increase in peak width as the temperature
is increased and the system becomes more disordered. Total loss of long-range
order is obvious at the transition temperature to the disordered
state.\cite{NovBruBav15}

As for the striped phase, while it is possible to generate many
stripes in the constrained geometry, it was not possible to generate more than
six to eight stripes using the unconstrained geometry and a system size of 20K.
The structure factor peaks for this case, for both constrained and
unconstrained geometries, possess strong satellites once there are more than a
couple of domain walls in the system.
The presentations of the experimental data for $S(\bm{Q})$ in
Refs.~[\onlinecite{KerDavZep87,Ker87}]
do not seem to show the existence of strong satellites as found in these
MD simulations.  It does appear that the existence of strong satellite peaks
is inconsistent with the data since the experimental analysis assumed
the main (parent) peaks are the major contributors to the scattering intensity.
The lack of strong satellite peaks is an indication that even if the
experimental system is a SIC phase, it is not one possessing many stripes.
The main peaks for the simulated (MD) striped phase did show
shifts from the \Rt{3} peak locations, although these shifts might not
have a simple relation to the ``misfit'' and the main satellite peak was
similar in intensity to the parent peak.
While our
comparisons of the MD results with the experimental data is more qualitative
then quantitative, these comparisons did take into consideration peak
locations, peak shapes, and peak intensities. We believe the conclusion
that the experimental diffraction peaks are not consistent with the SIC phase
of the BR model are based on sound arguments.

As would be expected, peaks for the apparently chaotic phase show a variety
of structures.  Many of these are difficult to interpret, but are
reflective of the disorder in the system.  Some peaks show similarities to
the peaks shown in the HAS data\cite{KerDavZep87} associated with
the SIC phase, even though the MD system is not a striped structure.
\begin{figure}
   \centering
   \includegraphics[width=2.8in]{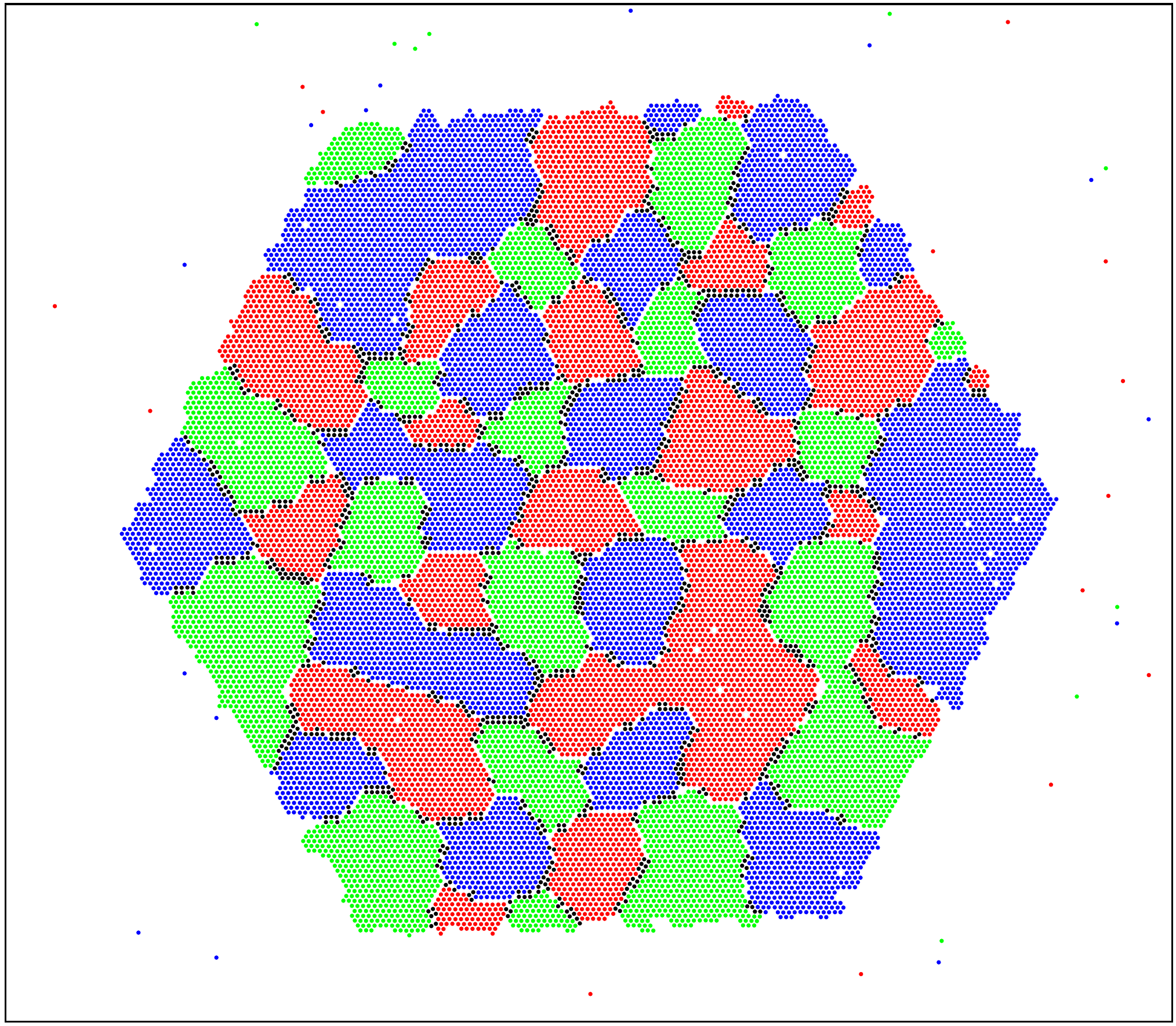}
   \caption{\label{Fig_Domains}(Color online) Domain structure for
    case study BR-H:20K.  The initial rotation is $2^{\circ}$ and the
    temperature is 2.922~K. Each shade (color) represents a domain of a
    different sublattice, while black (white) represents enhanced (reduced)
    density domain walls.
    See Ref.~[\onlinecite{BruNov08}] for details. }
\end{figure}
A good example of this is shown in Fig.~\ref{Fig_Domains}, where the domain
structure shown is an example of this chaotic phase. This figure shows the
typical pattern of domains and domain walls for these states, where the
definition of the three domain sublattices and the domain walls are to be
found in Refs.~[\onlinecite{BruNov08,NovBruBav15}].
This chaotic structure was generated by initialization in a low temperature
\Rt{3} patch configuration and then slightly rotating the patch (by about
$2^{\circ}$) before starting the simulation.  This system was first cooled
and then heated.  It is clear that this structure is not a striped phase, but
the structure factor, as shown in Fig.~\ref{Fig_SofQ_3D}, shows some
similarity to the experimental $S(\bm{Q})$ as shown in Fig.~2 of
Ref.~[\onlinecite{KerDavZep87}].  However, the $S(\bm{Q})$ peaks in the
simulation results are sharper than those found in the experimental results
and are also shifted in $\bm{Q}$-space.

\begin{figure}
   \centering
   \includegraphics[width=2.8in]{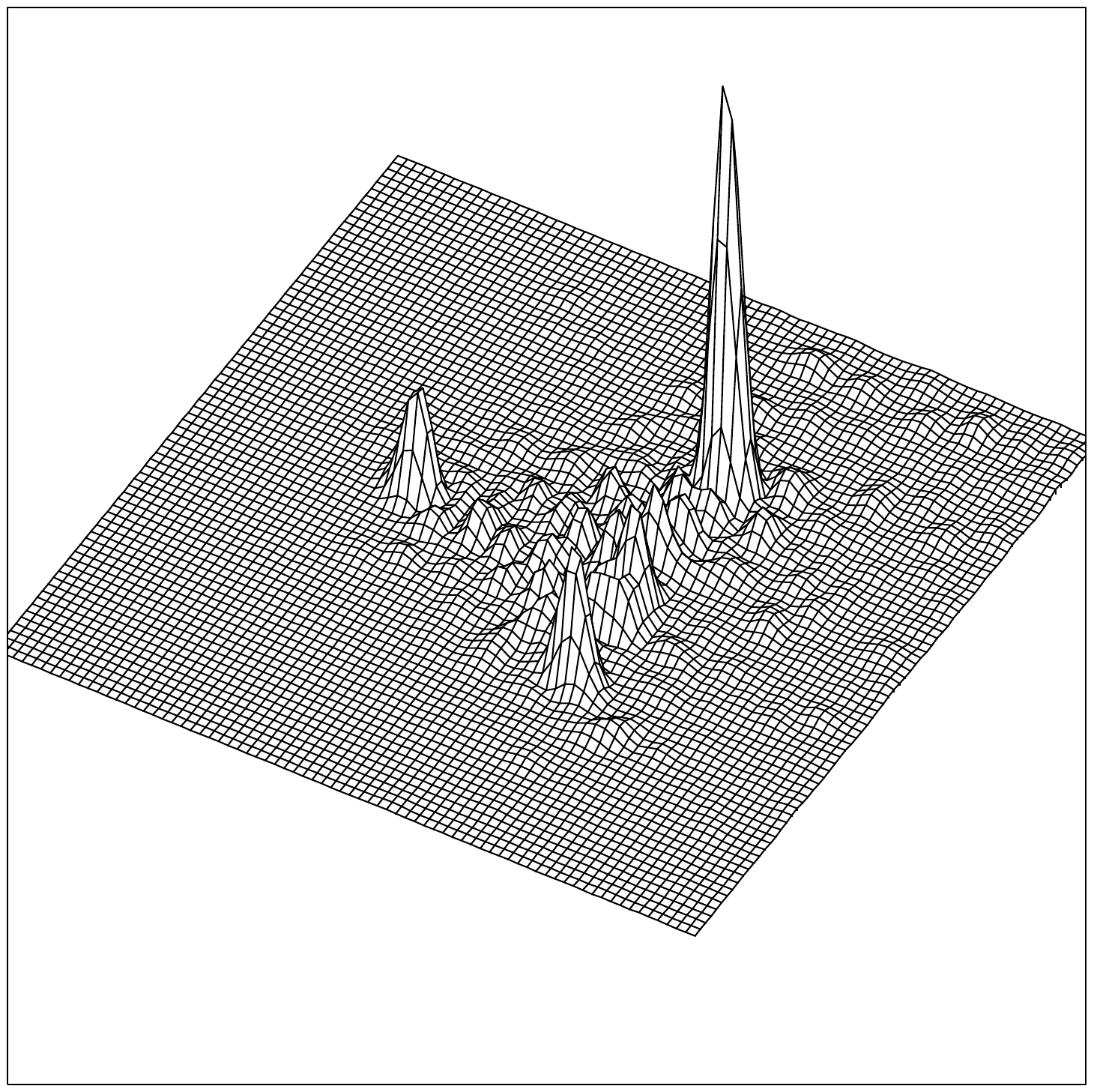}
   \caption{\label{Fig_SofQ_3D} 3D plot of $S(\bm{Q})$ for
    case study BR-H:20K.  The plot is for the region of $\bm{Q}$-space
    near the $(02)$ peak of the \Rt{3} xenon monolayer.  The origin is at
   the left corner, the $x$-axis is along the lower edge with
   $10.9 \le Q_x \le 11.9$, and the $y$-axis is along the left edge with
   $6.1 \le Q_y \le 7.1$.  The initial
    rotation is $2^{\circ}$ and the temperature is 2.922~K. The $\bm{Q}$
    values are normalized by the length-scale as discussed in
    Sec.~\ref{Model:MD}.  }
\end{figure}
\begin{figure}
   \centering
   \includegraphics[width=2.8in]{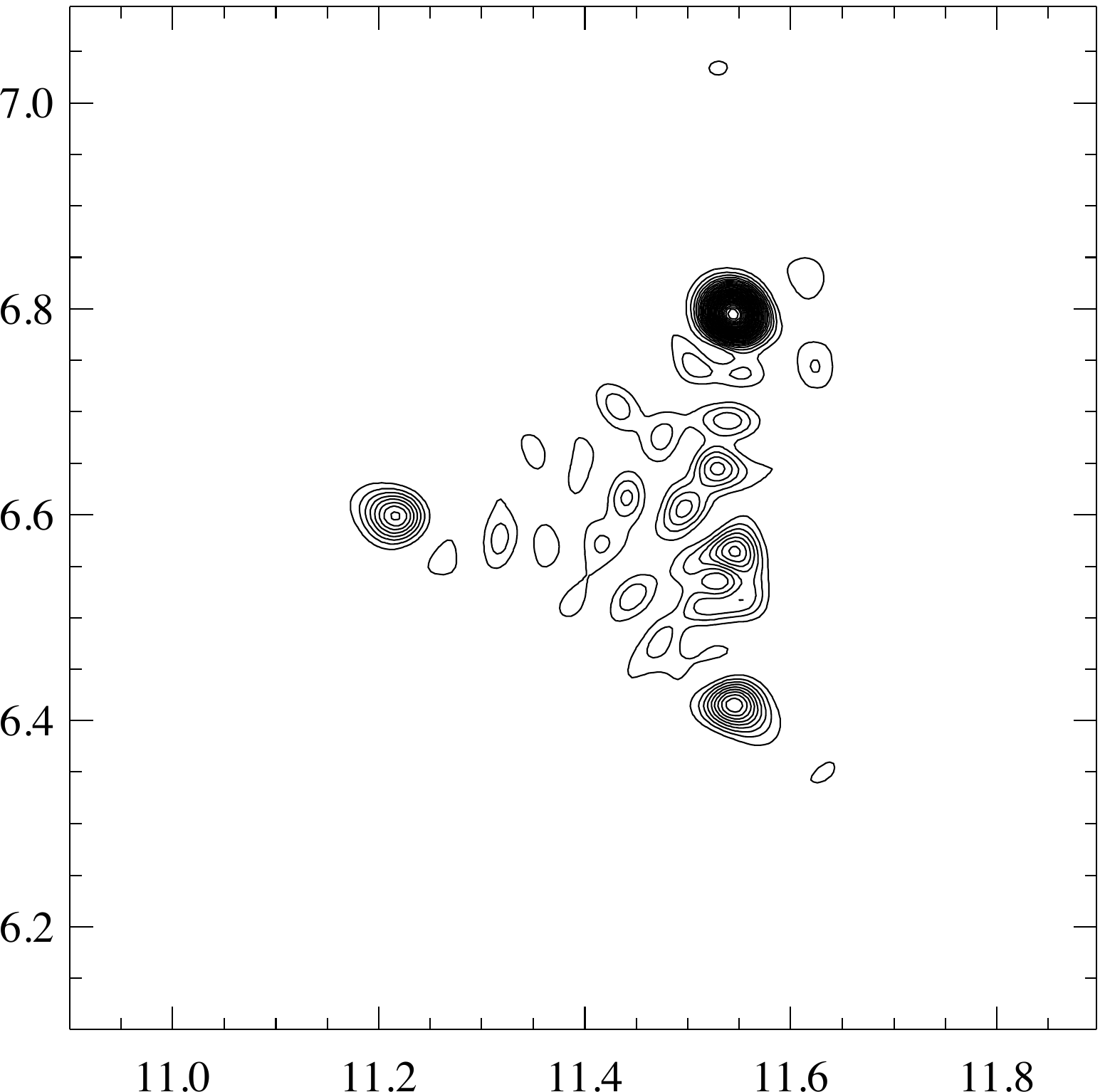}
   \caption{\label{Fig_SofQ_Contours} Contour plot of $S(\bm{Q})$ for
    case study BR-U:20K.  The plot is for the region of
    $\bm{Q}$-space near the $(02)$ peak of the \Rt{3} xenon
    monolayer.  The horizontal axis is $Q_x$ with
    $10.9 \le Q_x \le 11.9$, and the vertical axis is $Q_y$ with
    $6.1 \le Q_y \le 7.1$.  The initial rotation is $2^{\circ}$ and the
    temperature is 2.922~K.   The $\bm{Q}$ values are normalized by the
	length-scale as discussed in Sec.~\ref{Model:MD}.}
\end{figure}
\begin{figure}
   \centering
   \includegraphics[width=2.8in]{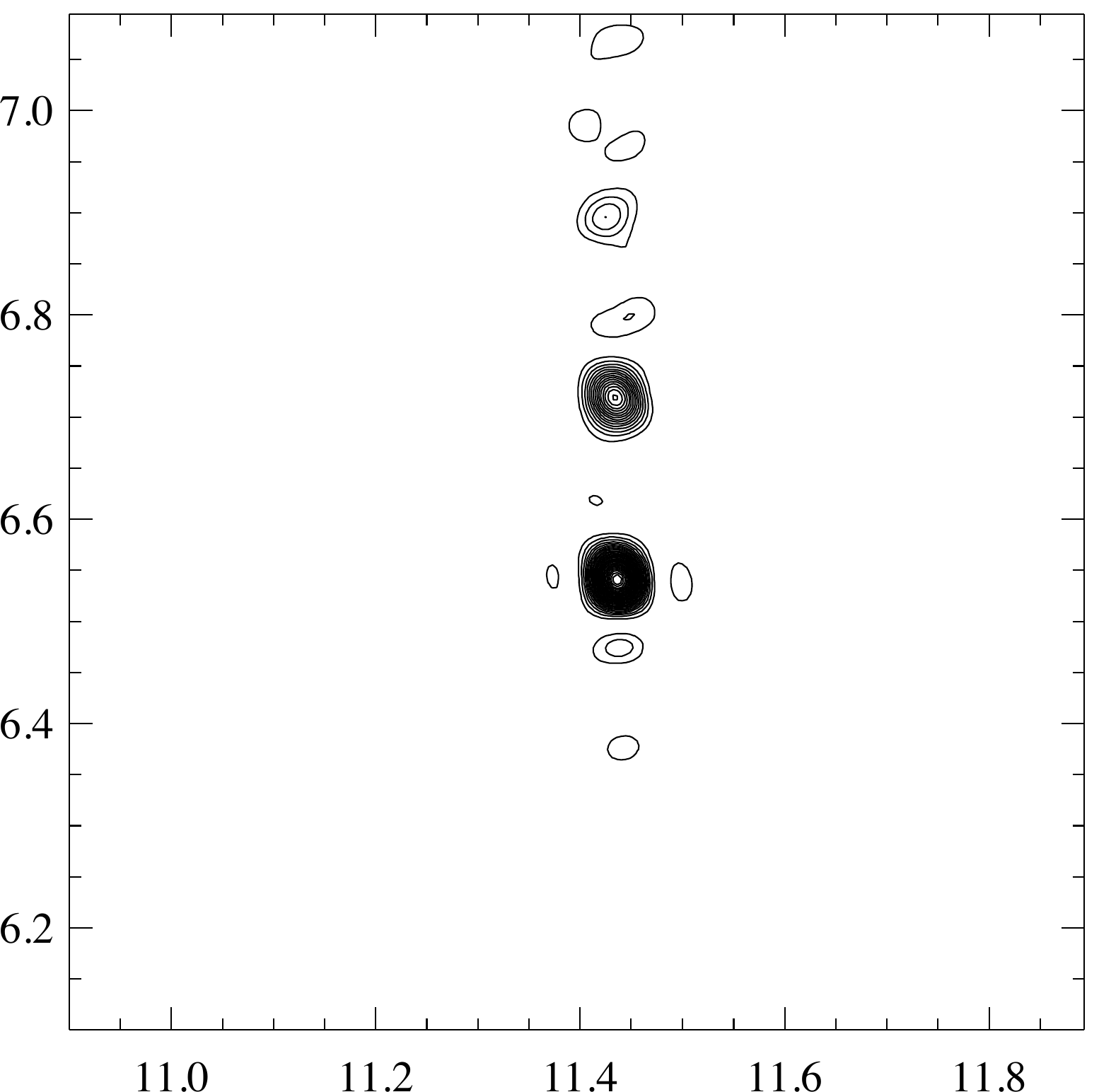}
   \caption{\label{Fig_SofQ_Striped} $S(\bm{Q})$ for
    case study BR-U:20K.  The plot is for the region of
    $\bm{Q}$-space near the $(02)$ peak of the \Rt{3} xenon
    monolayer.  The horizontal axis is $Q_x$ with $10.9 < Q_x < 11.9$, and the
    vertical axis is $Q_y$ with $6.1 < Q_y < 7.1$.  The initial configuration
    is striped and and the temperature is 31.56~K.   The $\bm{Q}$
    values are normalized by the length-scale as discussed in
    Sec.~\ref{Model:MD}.
    Note the high intensity of both the parent and main satellite peaks}.
\end{figure}
A contour plot of $S(\bm{Q})$ in the same region of $\bm{Q}$-space
is shown in Fig.~\ref{Fig_SofQ_Contours}. The largest peak is about two or
three times higher than the two smaller peaks, and the peak triplet is roughly
centered near the xenon $(02)$ peak of the \Rt{3} phase.
This can be compared to the $S(\bm{Q})$ that results from a striped
phase patch with five domain walls, as seen in
Fig.~\ref{Fig_SofQ_Striped}.  In the experiments, a triplet pattern was
explained as an incoherent sum of contributions from three wall orientations
rotated $120^\circ$ from the others, with the peaks being generated from the
shifted and unshifted parent peaks whose origins are the \Rt{3} peaks
equivalent to the xenon $(02)$ peak.  Adding together three rotated
contributions of the sort found in Fig.~\ref{Fig_SofQ_Striped} would
not give a triplet pattern.
%}

\section{\label{Comparison}Comparison to Experiment}

Connections with the experimental results were investigated by calculations
of the static structure factor $S(\bm{Q})$\cite{BruNov08,NovBru14} for a
selected subset of runs spaced along an appropriate range of temperatures.
These calculations were compared to the experimental results to see if there
are other possible interpretations of those experiments.  Since the BR
projection is the most appropriate one for the very low temperature range,
the focus was on that projection using an unconstrained geometry with $20$K
atoms in a single patch and having an average density of roughly
$0.5 \rho_0$. Results for the U35-H projection are essentially
the same as the BR-H projection.
We explored variations in the initialization of the system so as to generate a
variety of initial structures.  These variations used a series of initial
rotations and initial densities (using the unconstrained geometry) to produce
initial configurations of the SIC, AIC, and RIC structures.  In
addition, the response of the system to changes in the corrugation was
investigated.

Simulations of the low temperature submonolayer solid show a system
with extremely narrow domain walls that tend to zigzag in a rather haphazard
and perhaps chaotic fashion.\cite{Note-Stud}  The domain walls are often only
two or three atoms in width, the width varying along the length of the wall.
Some stretches of these walls exhibit wall widths which are effectively zero
(that is, domains of different sublattices directly abut each other
with a small gap).  This is in marked contrast to the walls found in the Xe/Gr
system, where the domain walls have a regular structure, are relatively wide,
and are essentially of constant width.\cite{BruNov08,NovBru14} Furthermore, the
walls in the Xe/Pt system appear to be more erratic and not as easily
categorized as domain wall models typically used in calculations found in the
literature.\cite{KerDavZep87,SpeMakPet87,KarBer82} In addition, the walls
seem to be
rather resistant to movement (as evidenced by their propensity to stay near
their original position as the system evolves). This happens both as the
running time is increased and as the temperature is raised. Furthermore, this
occurs even when a thermodynamic analysis clearly indicates that the state
in question is not the one with the most thermodynamically stable structure.
That is, these domain walls seem to stabilize meta-stable states, behaving as
if they are pinned at low temperatures.  As the temperature is raised above
60~K, these walls then appear to relax, causing the system to form a proper
\Rt{3} structure with a couple of large domains (although often surrounded by
some disorder as the temperature approaches the transition temperature).

The prediction of the BR model for the structure of the low temperature
submonolayer is in stark contrast to both the HAS results
and the STM results.\cite{Ker87,KerDavZep87,DieSeyCar04,BruSchFed02}
The existence of large patches of irregular but roughly hexagonal
\Rt{3} domains separated by very narrow and irregular domain walls not only
generates strong, single peaks at those scattering vectors $\bm{Q}$ that are
coincident with the reciprocal lattice vectors $\bm{G}$ of the substrate, but
produces the triplet pattern shown in Fig.~\ref{Fig_SofQ_3D} for those
$\bm{Q}$ not near such $\bm{G}$. However, the scattering
pattern for the BR model, even with thermal smoothing, does look different
from that shown in the experimental work of Kern.\cite{KerDavZep87}
Furthermore, these differences exist for both the ``chaotic hexagonal'' phase
and the ``chaotic striped'' phase.

If one compares the $S(\bm{Q})$ for the ``chaotic hexagonal'' phase of the
BR model to the experimental results, it is possible to see that the
peaks in the vicinity of those $\bm{Q}$ vectors that are near corresponding
reciprocal lattice vectors of the surface (for example,
$\bm{Q} \approx \bm{\tau}_{(\bar{1}2)} \approx \bm{G}_{(01)}$) are actually
not that different from the experimental results.  The calculations show
a strong, single peak at the location of the appropriate $\bm{G}$ vector
(that is at the corresponding \Rt{3} phase $\bm{\tau}$ vector).  The Kern
data shows a strong peak with (what is described as) a very weak and indistinct
doublet. Here, the data is really not that different from the MD simulation.
However, for the scattering peaks near the \Rt{3} phase $\bm{\tau}_{(02)}$
vector, the BR model shows a triplet centered about that location while the
Kern data shows a triplet displaced significantly outward from this location.
Furthermore, the peaks in the BR calculation are more distinct, possessing
significantly smaller widths, then those found in the experimental
results.\cite{Ker87,KerDavZep87,DieSeyCar04,BruSchFed02}

Now, if one would instead examine the ``chaotic striped'' phase, looking at
both the BR model and the Kern data,
comparing $S(\bm{Q})$ for the ``chaotic striped'' phase to the experimental
results, what is true of the ``chaotic hexagonal'' phase with regard to
scattering near the $\bm{Q} \approx \bm{G}$ vectors is also true for the
``chaotic striped'' phase.  Namely, the only significant intensity is at the
corresponding \Rt{3} parent peak.  However, for scattering near other peaks
(like $\bm{\tau}_{(02)}$) the BR ``chaotic striped'' phase would show the
pattern in Fig.~\ref{Fig_SofQ_Striped} added (incoherently) with two others
rotated by $\pm120^{\circ}$.  This appears to be inconsistent with the Kern
experimental results since, given the strong satellites shown in
Fig.~\ref{Fig_SofQ_Striped}, the calculated peaks would not form a
simple triplet pattern.

\begin{figure}
   \centering
   \includegraphics[width=2.8in]{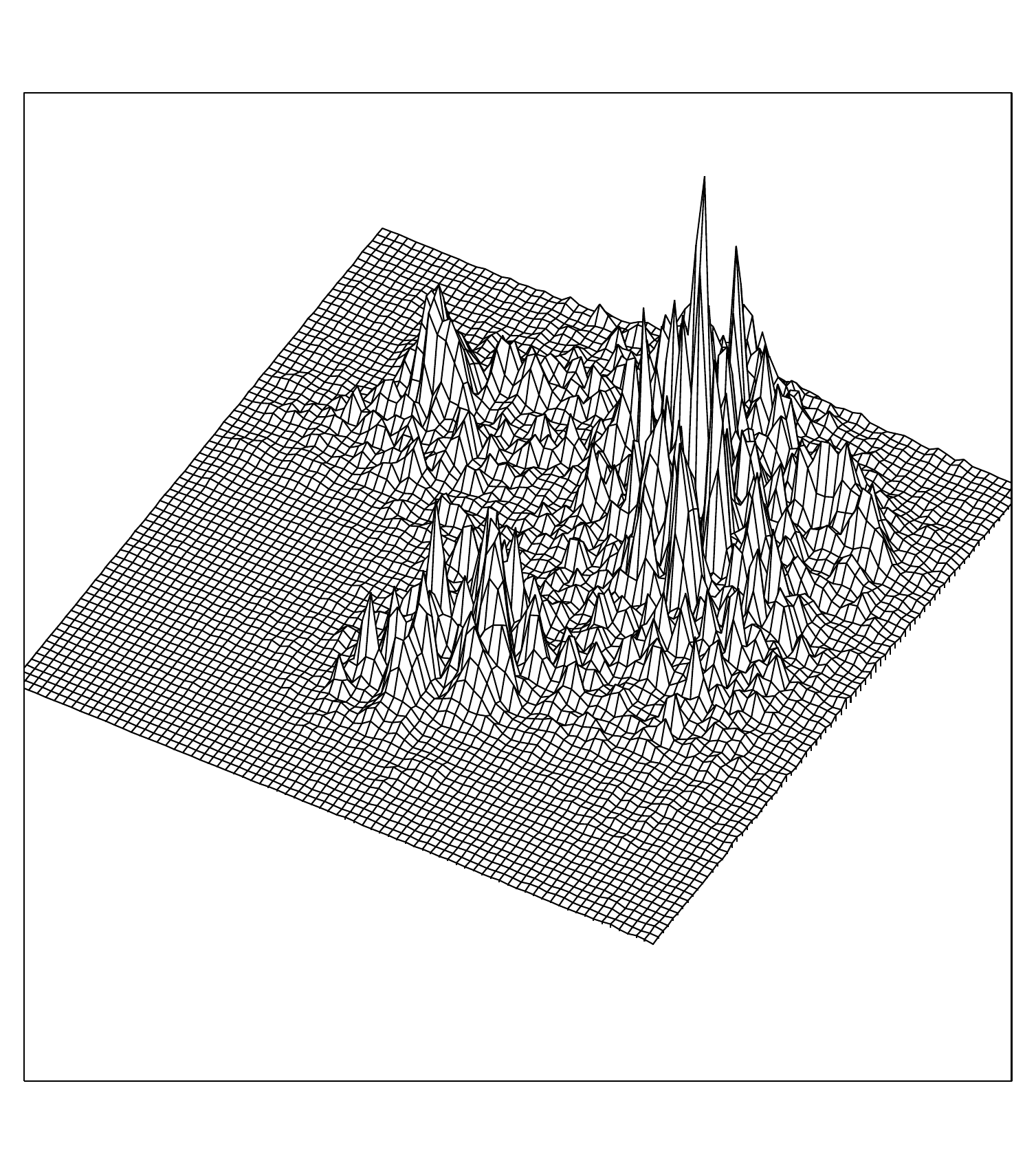}
   \caption{\label{Fig_Domains-II}(Color online) Domain structure for
    case study U20-H:20K.  The initial rotation is $2^{\circ}$ and the
    temperature is 45.26~K. Each shade (color) represents a domain of a
    different sublattice, while black (white) represents increased (reduced)
    density domain walls.
    See Ref.~[\onlinecite{BruNov08}] for details. }
\end{figure}

Since the predictions of the BR model, even with appropriate thermal smoothing,
appear to be at odds with the experimental results, questions about just how
the BR model could be deficient are relevant.  In
Ref.~[\onlinecite{NovBruBav15}], a similar problem arose when comparing the
experimental melting temperature and the prediction for the BR model.
This same problem arose in the determination of the mobility of a xenon atom
on the Pt$(111)$ surface by quasi-elastic helium atom
scattering.\cite{EllGraToe99} It does appear that there is a need
for further smoothing of the corrugation beyond the
thermal smoothing presented here. It can be suggested that this additional
smoothing might be due to thermal motion of the Pt$(111)$ surface
(see the supplementary material).                            % \cite{NovBavSM17}
But there also exists the possibility that the model corrugation is simply
too strong and (or) the well width in the surface-normal direction is too
narrow. Perhaps even at the very lowest temperatures, the basic BR model
parameters need tweaking.  With this in mind, we can examine the effects of
lowering the corrugation along the same lines as done in
Ref.~[\onlinecite{NovBruBav15}].

\begin{figure}
   \centering
   \includegraphics[width=2.8in]{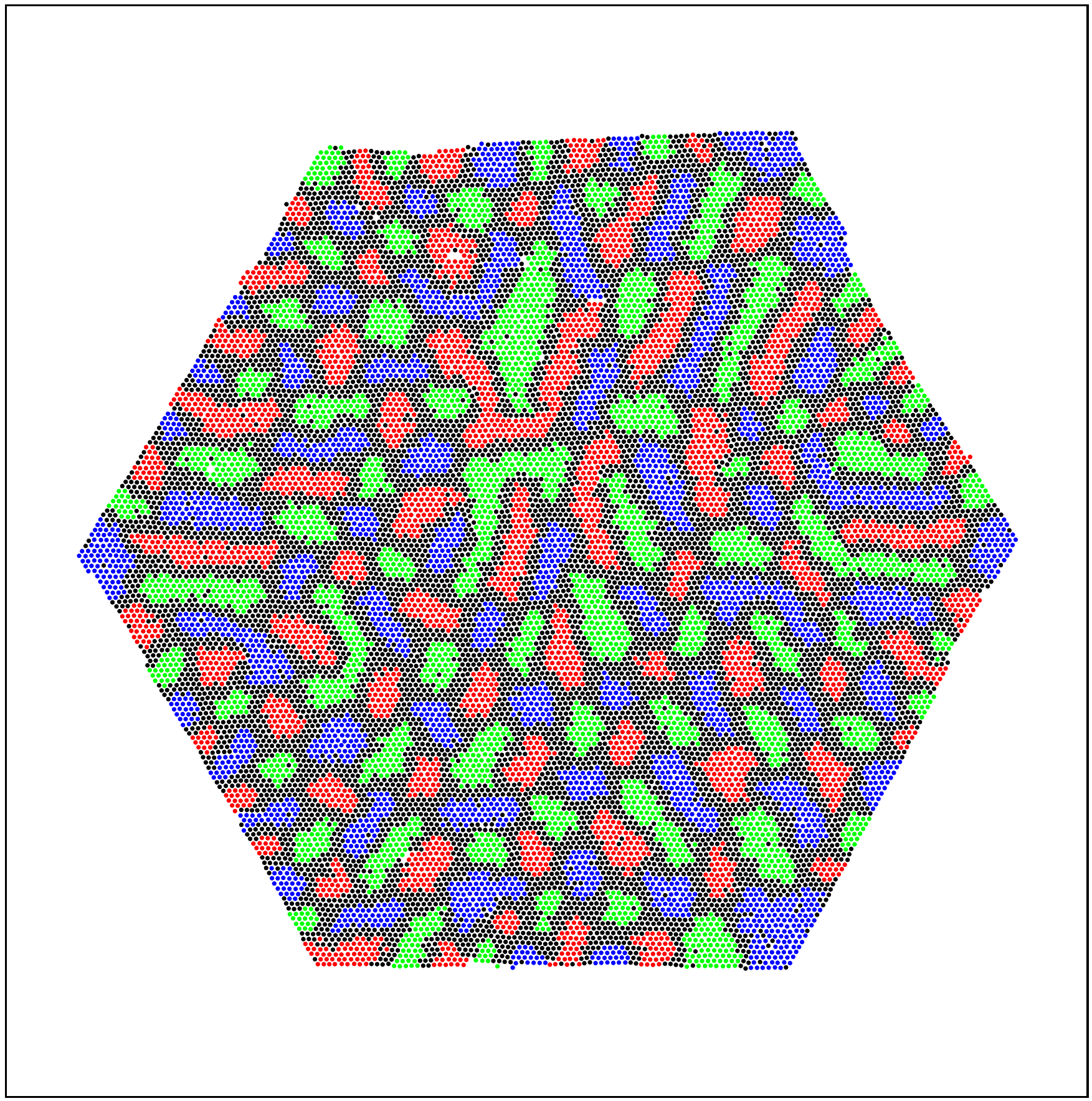}
   \caption{\label{Fig_SofQ_3D-2} 3D plot of $S(\bm{Q})$ for
    case study U20-H:20K.  The plot is for the region of $\bm{Q}$-space
    near the $(02)$ peak of the \Rt{3} xenon monolayer.  The origin is at
   the left corner, the $x$-axis is along the lower edge with
   $11.0 \le Q_x \le 13.0$, and the $y$-axis is along the left edge with
   $5.75 \le Q_y \le 7.75$.  The initial
    rotation is $2^{\circ}$ and the temperature is 26.4~K. The $\bm{Q}$
    values are normalized by the length-scale as discussed in
    Sec.~\ref{Model:MD}.  }
\end{figure}
For temperatures around 60~K, the BR value for $U_{(10)}$ is reduced (in
magnitude) by thermal (and quantum) smoothing to about 28 or 29~K as can be
seen in Table~\ref{Table:Q2D-QP}.
This reduction is not sufficient to alter the scattering pattern in any
significant way.
However, if the value of $U_{(10)}$ were in the range from about $ -25$ to
$ -20$~K, there is a significant shift in the position of the triplet
surrounding the $\bm{\tau}_{(02)}$ peak.
In particular, while there is still a triplet form
to the pattern, the center of that pattern moves ``outward'' and beyond the
\Rt{3} phase $\bm{\tau}_{(02)}$ position.  The spacial domain pattern is
similar to that shown in Fig.~\ref{Fig_Domains}, but the domains are smaller
and rather more irregular as can be seen directly in
Fig.~\ref{Fig_Domains-II}
(which is for the $U_{(10)} = -20$~K corrugation).

The $S(\bm{Q})$ for this state (at $T = 26.4$~K) is shown in
Fig.~\ref{Fig_SofQ_3D-2}.  The peaks in $S(\bm{G})$ are lower and less
distinct then in the BR case discussed above. Furthermore, the triplet is moved
outward in $\bm{Q}$ space, shifted relative to the xenon \Rt{3} $(02)$ peak
instead of being centered on that peak location, as it is in the BR case.
In fact, these peaks are quite similar to those in experimental
data.\cite{KerDavZep87} It must be noted that there
are many cases where the scattering pattern is different from those shown
here, often being more disorganized and without sharp peaks as would be
expected for a system with significant disorder. In particular, the state
of the system appears to be sensitive to the history of state formation.
This would be expected if the system is chaotic in nature,
but the existence of this sensitivity in a MD simulation is not proof
that the system is truly chaotic.

While structure factor calculations were not carried out for the ``chaotic
striped'' phase with $  | U_{(10)} |  < 30$~K, an analysis using a
different technique\cite{Nov79,Nov80,NovBru14} suggests that $S(\bm{Q})$
for such states would still produce noticeable satellites.  This would likely
conflict with the patterns observed in the Kern experiments as discussed
above.  It should be noted that reducing the corrugation so that
$U_{(10)} \approx  -20$~K does not eliminate the meta-stability and
``chaotic'' behavior.  However, doing so both increases the domain wall
width and decreases the temperature range of stability for the \Rt{3} phase.

\section{\label{Conclusions}Summary and conclusions}

Our molecular dynamics simulations of submonolayer Xe on Pt$(111)$,
using the Barker-Rettner model combined with the HFD-B2 Xe-Xe interaction
as modified by the McLachlan interaction, shows that the ground state of this
model is the \Rt{3} structure.  Furthermore, these BR model simulations
clearly show that the equilibrium low temperature, low pressure phase
is this same \Rt{3} phase. However, below approximately $60$~K, this
phase is susceptible to meta-stable chaotic disorder, creating domains of
various irregular shapes and producing structure factors similar in nature
to that of the HAS experiments.\cite{KerDavZep87} The interpretation of the
the HAS scattering as confirming the existence of a striped phase
as the low temperature structure may be a misinterpretation of the actual
situation.  Finally, these simulations show that this disorder anneals out
as the temperature is raised above the $60$~K mark, driving the chaotic
system back into the \Rt{3} structure. It is from this higher temperature
\Rt{3} phase that the system melts (disorders).\cite{NovBruBav15}

The susceptibility to chaotic behavior is exhibited by a mix of disorganized
hexagonal and striped domains in the same sample.  The resulting $S(\bm{Q})$
near the xenon $\{10\}$ and $\{20\}$ peaks shows a triplet pattern that is
similar in structure to that seen in
HAS experiments.\cite{KerDavPal86,Ker87,KerDavZep87}
On the other hand, for $S(\bm{Q})$ near the xenon $\{11\}$ peaks, the MD
analysis shows essentially no difference between reflections from any
of the examined phases (the \Rt{3}, the HIC, the SIC , and the chaotic
structures).  All phases show a strong peak at the platinum $\{10\}$
reflections,
namely just those found in the \Rt{3} phase. This is understandable
since all these structures have the vast majority of the xenon atoms at or
close to adsorption sites with, at most, only a few percent of these atoms
in very narrow and irregular placed domain walls. Given that the \Rt{3} xenon
$\{11\}$ peak set is coincident with the platinum $\{10\}$ peak set, all the
xenon atoms at adsorption sites reflect in phase with each other at these
values of $\bm{Q}$.  These results contradict statements in
Ref.~[\onlinecite{KerDavZep87}] about the analysis of the scattering
from the xenon $\{11\}$ peaks of the SIC structure, at least as
it could be interpreted for the submonolayer case. Furthermore, the simulations
of the model SIC phase show strong satellites in $S(\bm{Q})$, which
contradicts that same experimental analysis for the
$\{10\}$ and $\{20\}$ peaks.

There are, however, some problematic
issues associated with the BR model, especially at higher temperatures.
In particular, comparing simulation results to experimental results,
there seems to be more smoothing of the corrugation with increasing
temperature than can be accounted for within this model.
The effect of this reduced corrugation on the thermodynamic behavior is
examined in detail in Ref.~[\onlinecite{NovBruBav15}], but this behavior
is also found in quasi-elastic helium atom scattering (QHAS) experiments
which can examine the diffusion of a Xe adatom on the Pt(111)
surface.\cite{EllGraToe99}
The quasi-two-dimensional approach we used, combining a classical treatment
of the monolayer dynamics parallel to the surface
with a quantum treatment of the dynamics perpendicular
to the surface, does help mitigate these problems, but it does
not fully resolve all the issues. Our investigation of the thermodynamic and
structural
properties, comparing our calculations to previous simulations and known
experimental results of this and other systems, indicate that there is more
thermal smoothing near the melting temperature than can be accounted for by
current models.
This is in stark contrast to other systems, such as Xe/Gr, where the
same approach does an excellent job of explaining the experimental
results.\cite{NovBru14}

Further experimental studies are important for progress
in the understanding of this system.  In particular, work
that can better examine the very low temperature corrugation of this system
and probe the Xe/Pt potential energy surface is needed.
For example, a careful experimental study of single-particle diffusion
from very low temperatures up through melting,
combined with corresponding simulations which include quantum corrections at low
temperatures, could go a long way to
the determination of the corrugation and its dependence on temperature.
Improved ab~initio
studies of the xenon-platinum potential energy surface would be very useful,
although the precision needed may be beyond the limits of current theoretical
analysis.\cite{BetBir00}
Calculations of the effects of the thermal motion of the Pt$(111)$
surface on the behavior of the xenon monolayer could be critical to the
understanding of this problem.  It is also important to do an experimental
study of the effects of surface dynamics on the surface corrugation
as the temperature is raised.  It is possible that the
dynamics of the platinum surface significantly influences the dynamics of
the xenon monolayer at high temperatures by significantly smoothing the
surface corrugation.
Also important is the study of similar systems, namely heavy noble gases
adsorbed on strongly corrugated substrates having dilated adsorbate lattices.
One such example could be submonolayer xenon adsorbed on
Ru$(001)$.\cite{NarMen98}  Furthermore, the structural analysis of the
scattering from these systems must include an examination of
possible chaotic states of the sort observed in these simulations.

The BR model does a good job of explaining the low temperature
behavior of the xenon monolayer on the Pt$(111)$ surface. However,
while it is able to reproduce a significant collection of data, it is clear
that getting the transition temperature for melting right and explaining
the scattering data, the STM data, and the QHAS data requires
alterations in or enhancements to the BR model at high temperatures.
While reducing the
corrugation does effect the phonon spectrum,\cite{BruNov00}
a simple SCP calculation shows that reducing the corrugation from the
BR value to $U_{(10)} \approx -20$~K, reduces the in-plane zone-center
phonon gap by about 25~\% and produces a small increase in the maximum
in-plane phonon energy.\cite{Note-Novaco} However, these shifts look to
be borderline tolerable as to the maintaining of the agreement with
the previously calculated phonon energies and the experimental
data.\cite{BruNov00, BruGraToe00} For all its successes, and there are
many, the BR model seems to need improvements of the sort discussed here.

\section{\label{Supplementary}Supplementary Material}

See the supplementary material for additional justifications, explanations,
calculations, figures, and tables.  This material includes discussions of
1: Additional background information about this system;
2: Details of the simulation methodology;
3: Details of the free energy analysis;
4: Details on the calculations of the Q2D Fourier coefficients;
5: Calculations of the binding energies and heats of adsorption;
6: RMS vibrational analysis of the xenon monolayer;
7: Estimates of the effects of the platinum surface dynamics;
8: Additional results from and discussions of the simulations.

\begin{acknowledgments}
We would like to thank L.W. Bruch for probing discussions,
insightful questions, and useful suggestions.
We would like to acknowledge and thank C. Chen and S. Kapita for
the work they did on the preliminary studies that preceded and motivated
this work.  We thank Lafayette College for its generous support and the
Computer Science Department of Lafayette College for use of their
research computer cluster.
JB's exchange visit to Lafayette
College during the $2012$ calendar year was sponsored by the Brazilian
government agency
Coordena\c c\~ao de Aperfei\c coamento de Pessoal de N\'ivel
Superior (CAPES)
as part of the Science Without Borders program.  JB acknowledges funding
from the Austrian Science Fund (FWF) through the START project Y879-N27.
\end{acknowledgments}

\appendix

\section{\label{IndexMapping}Diffraction Peak Index Mappings}

\begin{table}
\caption{\label{Table:IndexMapping}
Equivalence mapping of the reciprocal lattice indexing used in other sources.}
\begin{ruledtabular}
\begin{tabular}{ l | c | c | c }
Lattice Type & Kern\footnotemark[1] & Maps Into & This Work \\
\hline
Platinum & $(\bar{1}\bar{1})$ & $\iff$ & $(10)$ \\
Xenon    & $(\bar{1}\bar{2})$ & $\iff$ & $(11)$ \\
Xenon    & $(\bar{1}\bar{1})$ & $\iff$ & $(10)$ \\
Xenon    & $(\bar{2}\bar{2})$ & $\iff$ & $(20)$ \\
\end{tabular}
\end{ruledtabular}
\footnotetext[1]{As described in Ref.~[\onlinecite{KerDavPal86f}].}
\end{table}
We are using the convention that the primitive translation vectors for both
the xenon lattice and the Pt$(111)$ lattice are placed $120^\circ$ apart. Thus,
the primitive reciprocal lattice vectors for both lattices ($\bm{\tau}$ for
the xenon and $\bm{G}$ vectors for the platinum) are $60^\circ$ apart. As a
result, the magnitude of $\bm{G}_{(11)} = \bm{G}_{(10)} + \bm{G}_{(01)}$ is
\Rt{3} times the magnitude of $\bm{G_}{(10)}$, and the same is true for
the $\bm{\tau}$ vectors. This is in contrast with some of the referenced
experimental work, where the opposite convention is used.  Furthermore, there
is a $30^\circ$ rotation and sometimes an inversion between the reciprocal
lattices used here and some of the experimental references.  In addition, the
SIC experimental data are an incoherent sum of peaks from three different
orientations of the SIC walls, while the $S(\bm{Q})$ peaks presented here are
those of a system with a single orientation.  Therefore, some care must be
exercised when comparing the simulation results to the experimental ones.
In particular, the convention used here for the reciprocal lattice vector
indexing differs from that used in the experimental work of
Kern and co-workers.\cite{KerDavPal86f} This work uses an angle of
$60^\circ$ between the primitive reciprocal lattice vectors and the
experimental work uses an angle of $120^\circ$. Table~\ref{Table:IndexMapping}
shows the mapping between the indices used in this work and that used in
Ref.~[\onlinecite{KerDavPal86f}].

% \twocolumngrid

\end{document}

\title{Supplement to: ``Simulations of submonolayer Xe on Pt$(111)$: \\
                        the case for a chaotic low temperature phase''}

\author{Anthony D. Novaco}\email[E-mail: ]{novacoad@lafayette.edu}
\affiliation{Department of Physics, Lafayette College \\ Easton,
                                                   Pennsylvania 18042, USA}

\author{Jessica Bavaresco}\email[E-mail: ]{jessica.bavaresco@oeaw.ac.at}
\affiliation{Institute for Quantum Optics and Quantum Information (IQOQI) \\
Austrian Academy of Sciences, Boltzmanngasse 3, A-1090 Vienna, Austria}

\date{\today}

\begin{abstract}
This supplement contains additional justifications, explanations, calculations,
figures, and tables to accompany the above paper by Novaco and Bavaresco.
\end{abstract}



\maketitle

\section{\label{Intro}Introduction}
 
We investigate whether chaotic behavior at low-temperatures can be present in
a case of physical adsorption; proposing that such is the case for Xe/Pt,
contrary to the common wisdom for this system.\cite{BruDieVen07}
This xenon monolayer exhibits very narrow domain walls with little distortion
in the ideal placement of the xenon atoms in the two domains on either
side of the wall; such behavior being driven by the strong corrugation and
the dilated adlayer lattice structure of this system.  It is this
characteristic that most likely generates the criteria needed to drive chaotic
behavior in this system.  The domain wall behavior in this system is consistent
with these walls being pinned, at low temperature, by the Peierls force; this
pinning force then being significantly reduced as the temperature
of the system is increased, causing the system to become more ordered.

The points just raised imply that a re-examination of this system is
appropriate. Given what we now know of the mutual interaction between Xe gas
atoms, the modification of this interaction by the substrate surface, and the
interaction between these atoms and the Pt$(111)$ surface, this re-examination
is both justified and timely.  Furthermore, today's computational
capabilities allow the simulation of much larger systems then what was
possible in the past.  In the case of Xe/Pt, this is important because some
of the more interesting effects shown here and elsewhere\cite{NovBruBav15}
would not appear in the small systems examined in the older works.

This supplement contains detailed discussions of the methodology of the
simulations, the explanation of the free energy analysis, details of
the calculations for the Q2D Fourier coefficients, calculations and
results for the RMS vibrational amplitudes of the xenon adatom, discussion
of the effects of platinum surface dynamics on the dynamics of the xenon
submonolayer, additional details of the results, and additional
discussions of these results.

\section{\label{MDmethod}Simulation Methodology}

The basics of the MD simulation are the same as those in
Refs.~[\onlinecite{NovBru14,BruNov08}], using essentially
the same code with mostly the same set of parameters.  As in those works, we
examine both the constrained geometry (a single phase filling the simulation
box) and the unconstrained geometry (an isolated patch surrounded by vapor).
Constrained-geometry simulations are used to determine the stable low
temperature phase for the classical system, and to examine the system in a
simple context (the constrained geometry having only one phase present at any
given temperature).\cite{NovBruBav15} These simulations of the constrained
geometry are also useful in the
interpretation of the high temperature behavior of the unconstrained-geometry
simulations.\cite{NovBruBav15} Simulations of the unconstrained-geometry are
carried out to examine both the thermal behavior and the structural properties
of the submonolayer patch.
The system sizes chosen were different from Ref.~[\onlinecite{NovBru14}], with
the largest system reported here being a constrained geometry with 65536
adatoms (65K).  Most of the simulations of the unconstrained geometry contained
20064 adatoms (20K).  Smaller systems having 4096 and 16384 adatoms (4K
and 16K) were also used to check for size dependencies.
%  in the experimental work.\cite{KerDavPal86}

The simulations were carried out using scaled equations with a length scale
of 4.3656~\AA, an energy scale of 282.8~K, and a time scale of 3.262~ps.
The scaling, time step, force truncation details, nearest-neighbor shells,
criteria for determining equilibrium, and the criteria for averaging are the
same as the earlier works cited.  The main difference in the behavior of these
MD simulations compared to those referenced above, is the existence of
significant meta-stability in these results that did not appear to occur in
the previous work (except for the largest corrugation reported in that work).
Relaxation times for the Xe/Pt simulations are much longer than those found
for Xe/Gr, which we attribute to the stronger corrugation in the Xe/Pt system.
The average densities for the unconstrained geometry varied over a wider range
in this study then in Ref.~[\onlinecite{NovBru14}].  Here the lower bound is
$0.14\rho_0$ and the upper bound is $0.67\rho_0$, with the typical simulation
carried out at $\approx 0.50\rho_0$.

The data was collected in blocks of 1000 time steps (32.6~ps), each time step
being 0.01 scaled time units.  In most cases, blocks 400--500
(13.04--16.30~ns) were used to calculate the thermodynamic functions as
before.  Some selective runs were carried out to nearly 1500 blocks
(48.90~ns).  Structure factor calculations typically used blocks 500--600
(16.30--19.56~ns), selecting about 25 to 50 configurations, uniformly spaced
in time, from that set of blocks.  The determination of the stable state is
obtained by initializations of the system in various structures and running
until the run has stabilized.  Equilibrium data was typically taken only
after this stabilization.  A given simulation was typically run for at
least 13.0~ns before being either heated or cooled to the next energy.  Each
run was then continued until equilibrium was attained.  The usual tests were
performed to determine if a given run has stabilized.\cite{BruNov00} 

The starting point for the \Rt{3}, AIC, and RIC unconstrained
configurations is a hexagonal patch centered in the simulation box.  The
initial patch density was adjusted by altering the lattice parameter of the
patch, and RIC configurations were generated by initial rotations of the
patch away from the \Rt{3} orientation.  The SIC initializations (for the
unconstrained systems) were rectangular patches with the \Rt{3} orientation,
commensurate with the platinum  structure in the xenon $[1,0]$
direction but compressed in the perpendicular direction.
The striped structures so generated typically had 2 to 8 domain walls.
Attempts to initiate unconstrained striped configurations with more than
8 walls result in systems
that are mechanically unstable and collapse into structures that are very
disordered and have higher free energy at a given temperature than
the \Rt{3} structure.  Attempts to initialize an unconstrained case with too
high a patch density also produces unstable runs that abort with a
corresponding of loss of energy conservation. The sensitivity
of the results to system size was tested by doing simulations of the
constrained geometry for 4K, 16K, and 65K sizes.  There is little
observable difference in the thermodynamics generated for those cases when
far from the melting transition, although this is not strictly true, as
explained in the Sec.~\ref{SuplConstrained},
for $\psi_0$ near the melting transition.

We did not study the approach to equilibrium near the melting
transition carefully, but we did carry out extensive testing of the
sensitivity of these results to the initialization of the system.  These
initializations involved changing the density, the orientation of the adatom
layer relative to the surface (the rotation angle), and the geometry of the
domain walls so generated (striped vs hexagonal). The relevant results
are described in Ref.~[\onlinecite{NovBruBav15}], in the main article,
and here in Sec.~\ref{SupplResults}.

To test the convergence of the $S(\bm{Q})$ results, calculations were
performed using the \Rt{3} structure, a 20K system, and configurations from
blocks 600 to 1100. These calculations show that, in general, it is only
necessary to use from 25 to 50 configurations, uniformly spaced over
about 100 blocks, to obtain good statistics. A larger sample set is not
necessary for a reliable $S(\bm{Q})$ calculation. Peaks for the
$(10)$, $(01)$, $(\bar{1}1)$, $(\bar{2}1)$, $(\bar{1}2)$, $(02)$, and
some others equivalent to these were generated and examined.  Calculations
were also carried out for selected \Rt{3} and striped structures of the
constrained geometry. The results were as expected.

\section{\label{FEdiscuss}Free energy analysis}

\begin{figure}
   \centering
   \includegraphics[width=2.8in]{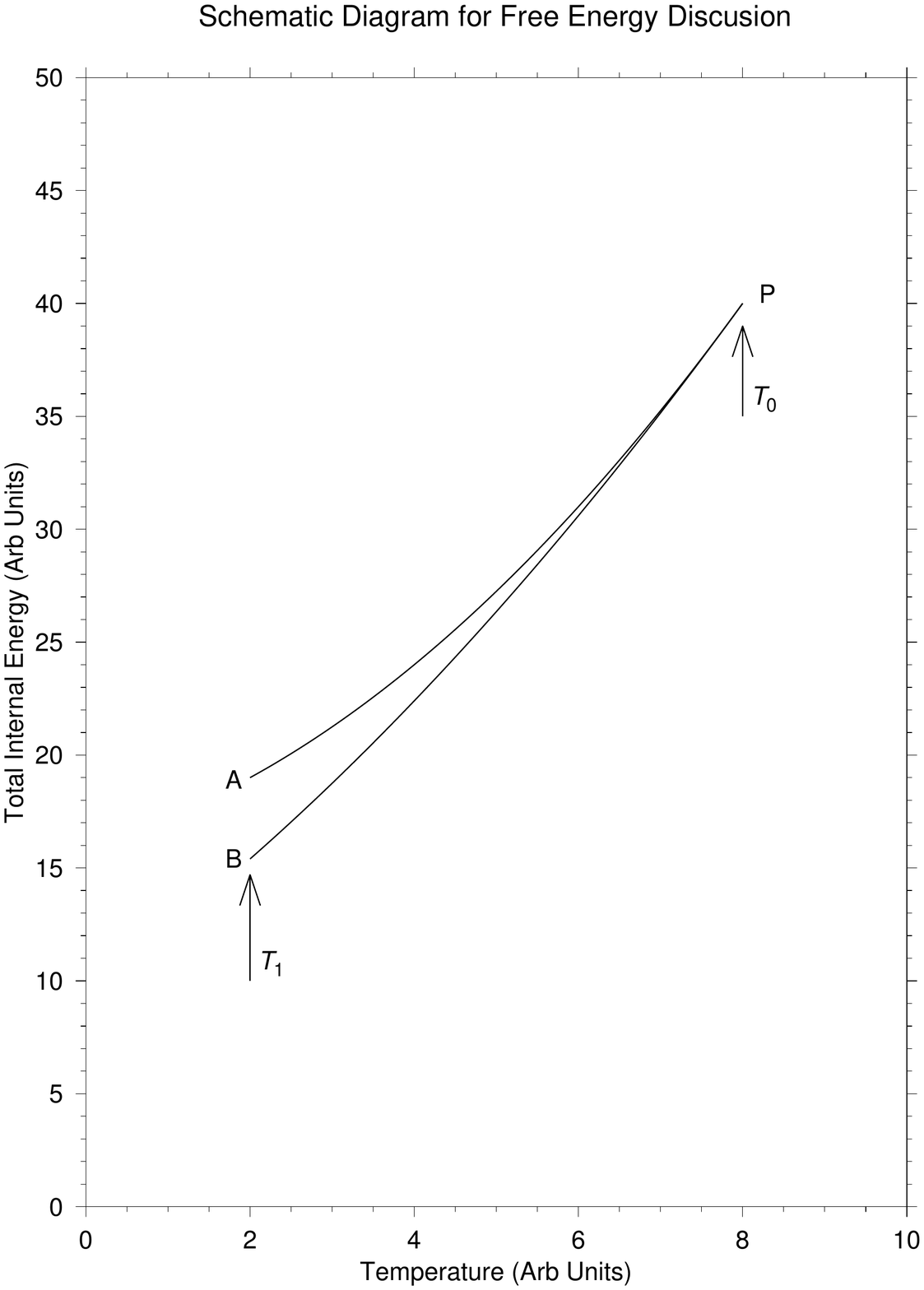}
   \caption{\label{FEpaths}Points $A$ and $B$, both at a temperature
    $T_1$, are terminal points for the corresponding paths.  Point $P$,
    the common endpoint for both paths, is at a temperature $T_0$.
    Both paths describe constant area processes along lines of
    presumed stable or meta-stable states.}
\end{figure}

Consider two paths in the energy-temperature plane, as shown in
Fig.~\ref{FEpaths}, that have a common point $P$ having a temperature of
$T_0$.  The combined path (A plus B)
describes a constant area process which both starts
and ends at points which have a temperature of $T_1$. Let path $A$ be above
path $B$, that is path $A$ has the higher energy at each temperature.
Now assume that each path represents a distinctly different structure
for the system, with each path being associated with a distinctly different
region of phase space. However, at the common point $P$, assume that
the two paths describe a common region of phase space, thus a common
phase of the system.  Assume that each path represents a meta-stable
system that, in the spirit of a restricted ensemble analysis, can be described
by equilibrium thermodynamics. Also, allow
for the possibility that one branch is actually the stable equilibrium
state of the system.  The question is: which of the two paths represents
the stable (or more stable) branch?

Consider a path which traverses path $B$ from $T_1$ to $T_0$, then moves
along path $A$ from $T_0$ to $T_1$.  Start with the identity
\begin{equation*}
f_A - f_B =  \left ( \epsilon_A - \epsilon_B \right )
                                  - T_1 \left ( s_A - s_B \right )
\end{equation*}
where $f$, $\epsilon$, and $s$ are the Helmholtz free energy, the internal
energy, and the entropy (each per particle).
We assign points $B$ and $A$ to represent the corresponding endpoints of
the path first along path $B$ through $P$ and then along path $A$,
Now, if $c_a$ is the specific heat (per particle) at constant area,
then along either path we have
\begin{equation*}
\epsilon_f - \epsilon_i = \int_{T_i}^{T_f}\diff{T} c_a
\quad \text{and} \quad
s_f - s_i = \int_{T_i}^{T_f}\diff{T} \frac{c_a}{T} .
\end{equation*}
Using the relations above, we generate a simple relation between the
free energy at point $A$ and that of point $B$, namely:
\begin{equation}
\label{FEequation}
f_A - f_B  = \int_{T_1}^{ T_{0} }\diff{T}
             \left [ 1.0 - \frac{T_1}{T} \right ]
                \left [ c_a^B - c_a^A \right ]  ,
\end{equation}
where the superscript on $c_a$ refers to the path taken.
The case shown in Fig.~\ref{FEpaths} illustrates the case most often
found in these simulations.  In particular, the path with the higher energy
also has the smaller specific heat (smaller slope) with the paths merging
at some higher temperature.
Thus in Eq.~(\ref{FEequation}), we have $T_1 \le T$  and $c_a^B > c_a^A$,
which implies that $f_A > f_B$ and the lower path is not only the one of
lower energy, it is also the one with the lower free energy. Conclusion: the
lower path is the more stable one when $T_1 < T_0$.

For the $T_1 > T_0$, imagine the paths in
Fig.~\ref{FEpaths} inverted about point $P$,
with path $B$ possessing the higher energy, but still possessing the
larger $c_a$.  Repeating the analysis for this case, the end result is
similar to Eq.~(\ref{FEequation}), but has a sign change.  The result is:
\begin{equation}
\label{FEequation_rev}
f_A - f_B  = \int_{T_0}^{ T_{1} }\diff{T}
             \left [ \frac{T_1}{T} - 1.0 \right ]
                \left [ c_a^B - c_a^A \right ]  .
\end{equation}
With $T_1 > T_0$ and $c_a^B > c_a^A$: $f_A > f_B$. Therefore, path $B$ is
still the stable path. That is, for $T_1 > T_0$, the upper path is now
the stable one.  These arguments are similar to those found in
Ref.~[\onlinecite{Rief1965}], starting on page 296.

\section{\label{FourierQ2D}Q2D Fourier Coefficients}

The Q2D approach starts with the Steele expansion of the potential energy of
an atom due to the surface of a crystalline substrate.\cite{Ste73,Ste74}
This potential energy, denoted by $U(  \bm{r}, z )$, can be
written as a Fourier series of the form:
\begin{equation}
\label{SteeleEq}
U (  \bm{r} , z ) =
       \sum_{ \bm{G} } U_{\bm{G}} ( z )  \exp ({ i \bm{G} \cdot  \bm{r}} ),
\end{equation}
where $\bm{G}$ is a reciprocal lattice vector of the two-dimensional
surface lattice (See Appendix~A in the main article % ~\ref{IndexMapping}
for the naming and indexing conventions used).
It should be noted that the model presented here does not depend upon the
$U_{\bm{G}}(z)$ having any particular form, only that each be
expressed in some manner that allows the calculation of the
corresponding derivatives with respect to the $z$-coordinate.
Therefore, it is possible to use this approach in conjunction with any
potential energy model for the adatom-substrate interaction and not just
for the site-site model that is used in this communication.

The Fourier components that define the corrugation of the surface are
dependent on $z$, so if the adatom were to move along the surface following
the path of minimum energy, it would move perpendicular to the surface
as it moves parallel to the surface.  If this is treated as a purely
classical problem, there are a couple of ways to proceed.
The simplest method, and the one often used,\cite{BruColZar97}
is to find the minimum barrier to translation
moving from one adsorption site to the next and ascribe this to the
simplest form of Eq.~(\ref{SteeleEq}):
\begin{equation}
\label{SteeleU0}
U(\bm{r}) =  U_{(10)} \sum_{\bm{G}}
                    \exp ( i \bm{G} \cdot \bm{r} ),
\end{equation}
where $U_{(10)}$ is a constant, and
the sum over $\bm{G}$ is now only over the six reciprocal lattice
vectors equivalent to $\bm{G}_{(10)}$.
A somewhat improved (but more complicated) approach, is to first
calculate (for hexagonal lattices)
four specific values from Eq.~(\ref{SteeleEq});
these being the minimum value of $U_0 (z)$ and the minimum potential energies
over the Atop site, the hollow site, and the bridge site.
These energies are denoted by $U_0$, $U_A$, $U_H$ and $U_B$ respectively
with the corresponding $z$-value denoted by a corresponding $z_{opt}$.
From these four energies, the three independent Fourier coefficients
$U_{(10)}$, $U_{(11)}$, and $U_{(20)}$ can be obtained if it is
assumed that the higher-order coefficients can be ignored.  For the simple
hexagonal lattice (i.e. Pt$(111)$, we find:
\begin{subequations}
\label{UgEqs}
\begin{eqnarray}
U_{(10)} =&  5 \delta U_A /72 - \delta U_B /8 - \delta U_H /9  \\
U_{(11)} =&  ( \delta U_A + 2 \delta U_H )/18 \\
U_{(20)} =&  ( \delta U_A + 3 \delta U_B )/24
\end{eqnarray}
\end{subequations}
where $\delta U_A \equiv U_A - U_0$, $\delta U_B \equiv U_B - U_0$, and
$\delta U_H \equiv U_H - U_0$.  Similar equations hold for the
open-hexagonal-net lattice (i.e.  graphite) with the exchange of
$\delta U_A$ with $\delta U_H$.  For Xe/Pt, direct calculations show that
the second and third coefficients are much smaller than the first.

A third approach is a perturbation calculation for the displacements of
the xenon atoms in the $z$-direction, using a Fourier series expansion for
the optimum $z$-position and a Taylor series expansion of
the $U_{\bm{G}}(z)$ to second order.  It is then possible to define effective
2D Fourier coefficients $\tilde{U}_{\bm{G}}$ and easily show that:
\begin{subequations}
\label{Q2D_Eqs_Linear}
\begin{equation}
\sum_{\bm{G}_1}
            U^{zz}_{\bm{G} - \bm{G}_1 } ( z_0 )
                Z_{\bm{G}_1} = - U^{z}_{\bm{G}} ( z_0 )
\end{equation}
and
\begin{equation}
 \tilde{U}_{\bm{G}}  \equiv  U_{\bm{G}} ( z_0 ) +
            \frac{1}{2} \sum_{\bm{G}_1}
                  U^{z}_{\bm{G} - \bm{G}_1 } ( z_0 )
                         Z_{\bm{G}_1}
\end{equation}
\end{subequations}
where the $Z_{\bm{G}}$ are the Fourier amplitudes of the displacements in the
$z$-direction relative to $z_0$ and the $U^{z}_{\bm{G}}$ and $U^{zz}_{\bm{G}}$
are the first and second derivatives of the $U_{\bm{G}}(z)$. The expansion point
$z_0$ is the $z$-position of the minimum in $U_{0}(z)$.

The quantum projections corresponding to the first two methods use
similar ideas to the corresponding classical projections, but assume that the
$z$-dynamics is describable by wave functions that are products of single
particle wave functions in $z$, each with some $z_{opt}$ for its central
positions and having widths that depend upon the location of $\Psi_{\perp}$.
This approach treats the dynamics perpendicular to the surface with a SCP
approximation having (nearly) flat (constant frequency)
phonon modes polarized perpendicular to the surface and uncoupled to the
in-plane phonon modes.\cite{BruNov00} The parameters that define the
corresponding positional Gaussian distributions are determined from the
minimization of the contribution they make to the total energy (at zero
temperature) or the free energy (at finite temperature) of the system.
The potential energy terms in Eqs.~(\ref{UgEqs}) are replaced by the
corresponding averages of these potential energies over the corresponding
$\Psi_{\perp}$ Gaussians.  The $z$-positions and widths of these
Gaussians are determined self-consistently, in the spirit of standard SCP
calculations, as described and implemented in Ref.~[\onlinecite{Nov92}].
Thus the quantum treatment of the energies in Eqs.~(\ref{UgEqs}) involved
the average over both $z$ and $\bm{r}$ coordinates of these terms as described
in Ref.~[\onlinecite{Nov92}].

\section{\label{Heats}Binding Energies and Heats of Adsorption}

Connecting the heats of adsorption to the binding energy of a adatom to a given
substrate requires a model for the thermodynamics of the monolayer. Here we use
two rather simple but quite adequate models.  The connection being used between
the isosteric heat of adsorption $q_{st}$ and the theoretical binding energy
$\epsilon_0$ is either:
\begin{subequations}
\label{Q_st}
\begin{equation}
  q_{st} =  \frac{3}{2} kT + \epsilon_0
\end{equation}
or
\begin{equation}
	q_{st} =  \frac{5}{2} kT + \epsilon_0
\end{equation}
\end{subequations}
where $T$ is the absolute temperature.
The first equation is obtained by using the standard connection between
$q_{st}$ and the chemical potentials of both 2D and 3D ideal gases
(see Ref.~[\onlinecite{BruColZar97}], page 251 and
Ref.~[\onlinecite{Bai99}], page 157).  For this model, the assumption is that
the energy consists of the binding energy of the adatom to a flat surface plus
the kinetic energy contribution from an ideal, classical 2D gas.  Thus, the
binding energy $\epsilon_0$ is just $ - E_z$, the negative of the energy
associated with the $z$-coordinate using only the $U_0$ term.  Proper
inclusion of the effects of the corrugation should increase the binding energy
and isosteric heat estimates. The second equation is that of a 3D ideal gas
in contact with 2D lattice gas of 3D oscillators placed on random surface
adsorption sites.  Here, $\epsilon_0$ is the SCP energy of isolated, 3D
Einstein oscillators subjected to the full BR model interaction
(but no xenon-xenon interactions).\cite{Nov92,BruNov00} For this
model, the binding energy includes contributions from all three energy terms:
$E_z$, $E_{xy}$, and $E_{xyz}$.  These are calculated using a SCP treatment of
$N$ identical, isolated, anisotropic, 3D Einstein oscillators.  For
both theoretical models, the inclusion of the mutual interaction between the
xenon adatoms should increase both $\epsilon_0$ and $q_{st}$.  While these are
simple physical models, they should give acceptable estimates of the monolayer
thermodynamics.

\section{\label{VibsRMS}Xenon RMS Vibrational Amplitudes}

One approach to the determination of the corrugation that was explored
is the calculation of the RMS amplitude of the in-plane vibration of the
\Rt{3} phase.  This value has been determined
experimentally,\cite{DieSeyCar04} and can easily be calculated from
theoretical models. The calculation based upon the MD simulation uses the
same analysis as a calculation of the diffusion constant, mainly
calculating the average (over $t_0$ and all atoms) for the mean squared
displacement:
\begin{equation}
\label{diffusion}
\left < ( \bm{r}(t_0 + \delta t) - \bm{r}(t_0) )  \cdot
( \bm{r}(t_0 + \delta t) - \bm{r}(t_0) ) \right >  ,
\end{equation}
and examining this expression as a function of the time shift $\delta t$.
If this expression saturates as a function of $\delta t$, then there is no
diffusion and the saturation value (obtained by an average over $\delta t$
for large $\delta t$ values) is related to the RMS vibrational
amplitude.  In particular, if $\delta \bm{r}$ is the 2D displacement from
the equilibrium point and the dynamics of $\bm{r}$ is that of
a 2D isotropic oscillator with random initial positions,
expression (\ref{diffusion}) is equal to
$ 2 \left < \delta \bm{r} \cdot \delta \bm{r} \right > $,
or four times the square of the RMS amplitude of vibration
However, if there is diffusion, then the quantity in
expression (\ref{diffusion}) has a asymptotic linear dependence on
$\delta t$ with a slope that is proportional to the diffusion constant.

\begin{table}
	\caption{\label{DeltaRMS} Parallel and perpendicular RMS amplitudes
	of vibration for \Rt{3} xenon on the Pt surface.
	The temperatures $T$ are in kelvin and the vibrational amplitudes are
	in \AA.  }
\begin{ruledtabular}
\begin{tabular}{ l | c | c | c }
Source & $T$ & Parallel & Perpendicular \\
\hline
Diehl\footnotemark[1] & 80 & 0.23  & 0.17 \\
	& 110 & 0.35  & 0.17 \\
\hline
MD\footnotemark[2]& 29 & 0.17 & --- \\
	& 46 & 0.21 & --- \\
	& 88 & 0.29 & --- \\
	& 101 & 0.31 & --- \\
\hline
MD\footnotemark[3]& 33 & 0.20 & --- \\
	& 52 & 0.25 & --- \\
\hline
Q2D\footnotemark[4]&  80 & 0.25 & 0.13 \\
	& 110 & 0.29 & 0.16 \\
\end{tabular}
\end{ruledtabular}
\footnotetext[1]{Experiment: See Ref.~[\onlinecite{DieSeyCar04}].}
\footnotetext[2]{This work: Case study BR:64K.}
\footnotetext[3]{This work: Case study U25-H:20K.}
\footnotetext[4]{SCP theory using the full BR model as in
	Refs.~[\onlinecite{BruNov00,Nov92}].}
\end{table}

For the BR:64K case study, diffusion is absent
for $T \leq 101$~K.  However, for the U25-H:20K case study, diffusion starts
to be a problem at a temperature $T \approx 50$~K, and thus the estimated RMS
amplitude near that temperature is suspect.  Unfortunately, in this system,
both the MD and the SCP\cite{BruNov00} results indicate that the in-plane RMS
amplitude is relatively insensitive to the corrugation. Given that this
RMS amplitude was difficult to determine with high precision in the LEED
experiment,\cite{DieSeyCar04} it was not possible to use this approach to
narrow the range of possible values for the corrugation.  The experimental and
theoretical values are listed in Table~\ref{DeltaRMS}, and the agreement is
within the experimental uncertainty.  We interpret the MD and SCP results
to be an indication that the in-plane vibrational amplitude is dominated more
by the xenon-xenon interaction and less by the surface corrugation.  The
theoretical perpendicular values are calculated by the quantum approach
described in the main article.

\section{\label{PlatinumSurface}Effects of Platinum Dynamics}

The main article discusses mechanisms that could reduce the
corrugation beyond that found with the projection models presented here.
One such mechanism is the thermal motion of the platinum surface, which
would reasonably be expected to be more important at higher temperatures since
higher temperatures would imply larger amplitude oscillations of the surface.
A very simple estimate of the order of magnitude of such an effect can be
made by estimating the size of the surface normal motion of the platinum
and using that to estimate the possible variation in the corrugation.
The surface modes of platinum with a xenon monolayer have been examined, and
these modes show a strong, sharp peak in the platinum response (associated with
the surface normal motion) at a frequency such that
$\hbar \omega = 185.6$~K.\cite{BouBru04}
If this is used in a simple model, the one-dimensional SHO, then the
RMS deviation in the $z$-position of the platinum surface is given by the
equation:
\begin{equation}
  \delta z_{rms}^2 = \frac{\hbar}{2 M \omega} \coth ( 0.5 \beta \hbar \omega )
\end{equation}
where $M$ is the platinum atomic mass, $\beta$ is the inverse temperature,
and $\omega$ is the platinum response frequency.  For a temperature of
$10$ kelvin, this estimate gives $\delta z_{rms} = 0.026$~\AA~while at
$120$ kelvin it gives $0.032$~\AA. For comparison, the shift
in $z_{opt}$ as the xenon atom moves laterally over the platinum surface
barrier is about $0.10$~\AA, larger but not orders-of-magnitude larger than
the $\delta z_{rms}$ at $T = 120$~K.  Perhaps more relevant, the increase in
the $z_{opt}$ for a xenon adatom at the Atop site, as the temperature is
raised from 0 to 100~K, is about 0.05~\AA, and this produces a 20~\%
decrease in the effective corrugation of the surface.  Finally, simply
recalculating the $U_{\bm G}$ after increasing $z_{opt}$ by 0.04~\AA~ reduces
the corrugation by about 7~\%.  While the $\delta z_{opt}$ shift is small,
its effects might not be completely negligible at the transition temperature
as judged by comparison to other smoothing effects.  Furthermore, motion of
the Pt surface could prevent the xenon atoms from maintaining the optimal
$z$-distance from that surface, thus further reducing the effective
corrugation.  It is not unreasonable to expect further smoothing of the
corrugation due to this effect, but how much is not possible to determine
without examining the dynamical coupling between the adlayer and the substrate.
A robust argument can be made that this dynamical coupling could play an
important role in determining the melting temperature,\cite{NovMcT79} but this
needs more investigation using both theory and experiment.

\section{\label{SupplResults}Additional Results and Discussion}

Below about 60~K, different initializations did produce different
final results, often generating distinctly different structures
and causing small shifts in the thermodynamic functions.  
However, these low temperature runs appear to merge (as they were heated
above approximately 60~K) into a common set of thermodynamic values with
similar (although not identical) structures.  The exact
temperature at which this happens does vary from case study to case study.
Cooling from a state of uniform gas produced the same hysteresis as found in
Ref.~[\onlinecite{NovBru14}].

\subsection{\label{SuplConstrained}Constrained Geometry}

\begin{figure}
   \centering
   \includegraphics[width=2.8in]{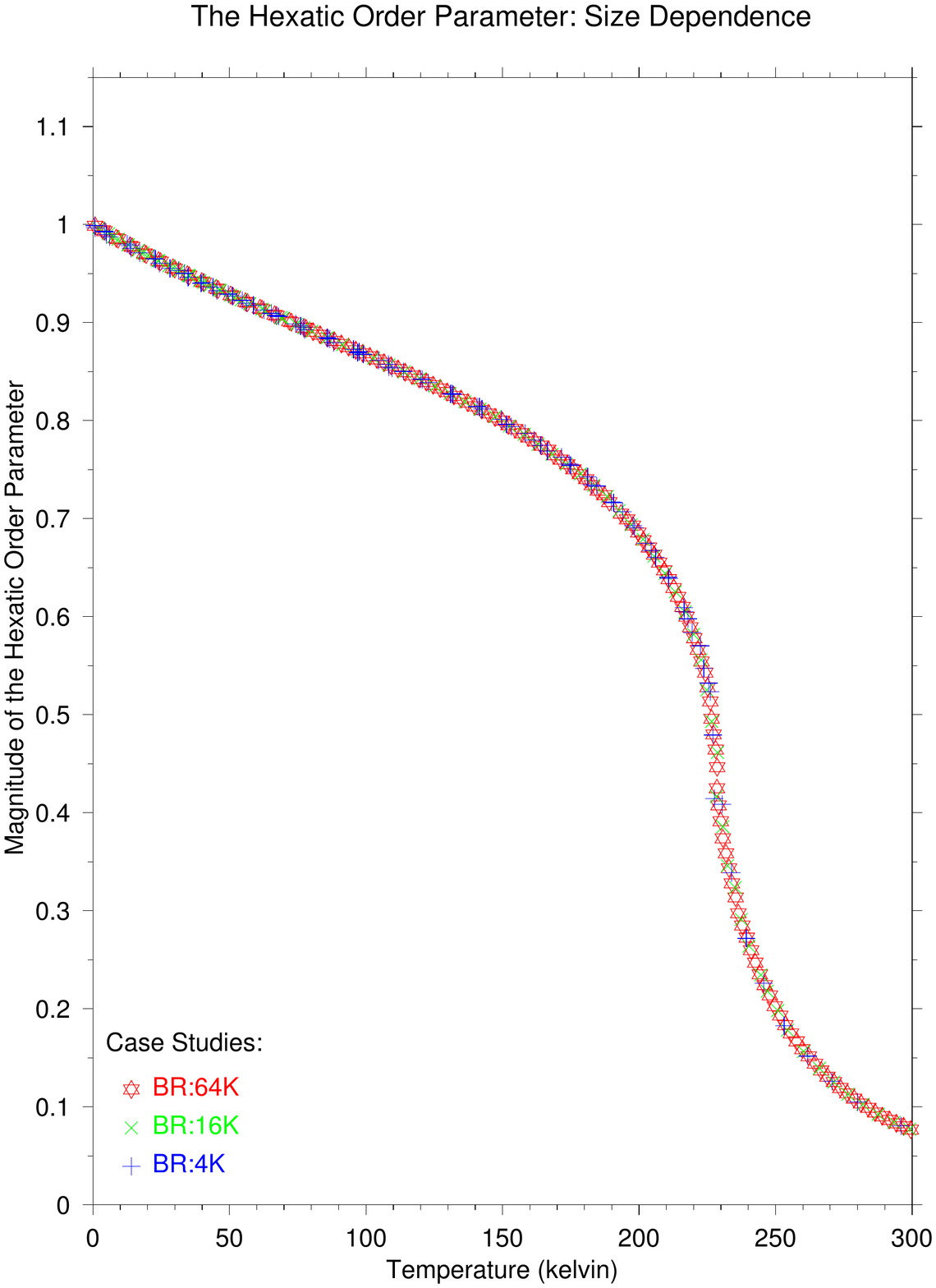}
   \caption{\label{NHpsi6.vs.Temperature:SizeDep} The
    hexatic order parameter of three case studies,
    illustrating the size dependence for constrained geometries.}
\end{figure}

Results for system sizes $64$K, $16$K, and $4$K show essentially the same total
energies at corresponding temperatures.  The differences (which are of
the order of $0.1$ kelvin) are well within the statistical noise. The
same is true for $\psi_6$, with any variation due to system size being within
the noise of the data as shown in Fig.~\ref{NHpsi6.vs.Temperature:SizeDep}.

\begin{figure}
   \centering
   \includegraphics[width=2.8in]{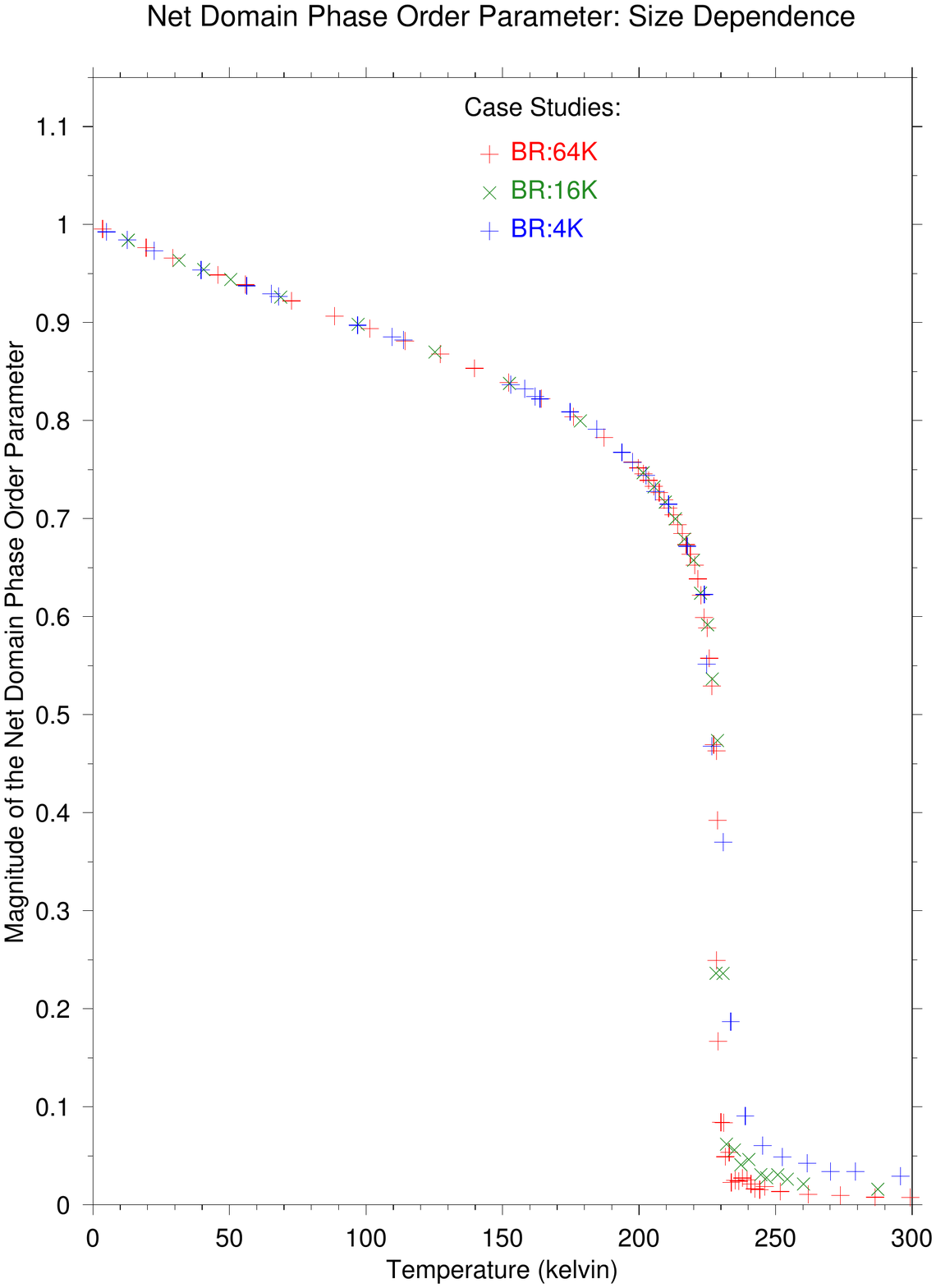}
   \caption{\label{NDPop.vs.Temperature:SizeDep} The NDP order parameter
    $\psi_0$ of three case studies, illustrating the size dependence for
    constrained geometries. The initialization is the \Rt{3} phase for all.}
\end{figure}

The drop in $\psi_0$ as the transition is approached from below, as seen in
Fig.~\ref{NDPop.vs.Temperature:SizeDep}, is far sharper than that for
$\psi_6$. This is consistent with $\psi_0$ likely being the more appropriate
order parameter to describe this order-disorder
transition.\cite{NovBruBav15} Furthermore, $\psi_0$ does show a variation
with system size in the tail of the curve just above the transition, as
illustrated by Fig.~\ref{NDPop.vs.Temperature:SizeDep}.
Postulating that $\psi_0$ is the appropriate order parameter for this system,
this sensitivity is to be expected.\cite{NovBruBav15}  The insensitivity
of $\psi_6$ to system size, and its more gradual
drop to zero with increasing temperature, implies that it is not likely the
most appropriate order parameter for this transition.\cite{NovBruBav15}

\begin{figure}
   \centering
   \includegraphics[width=2.8in]{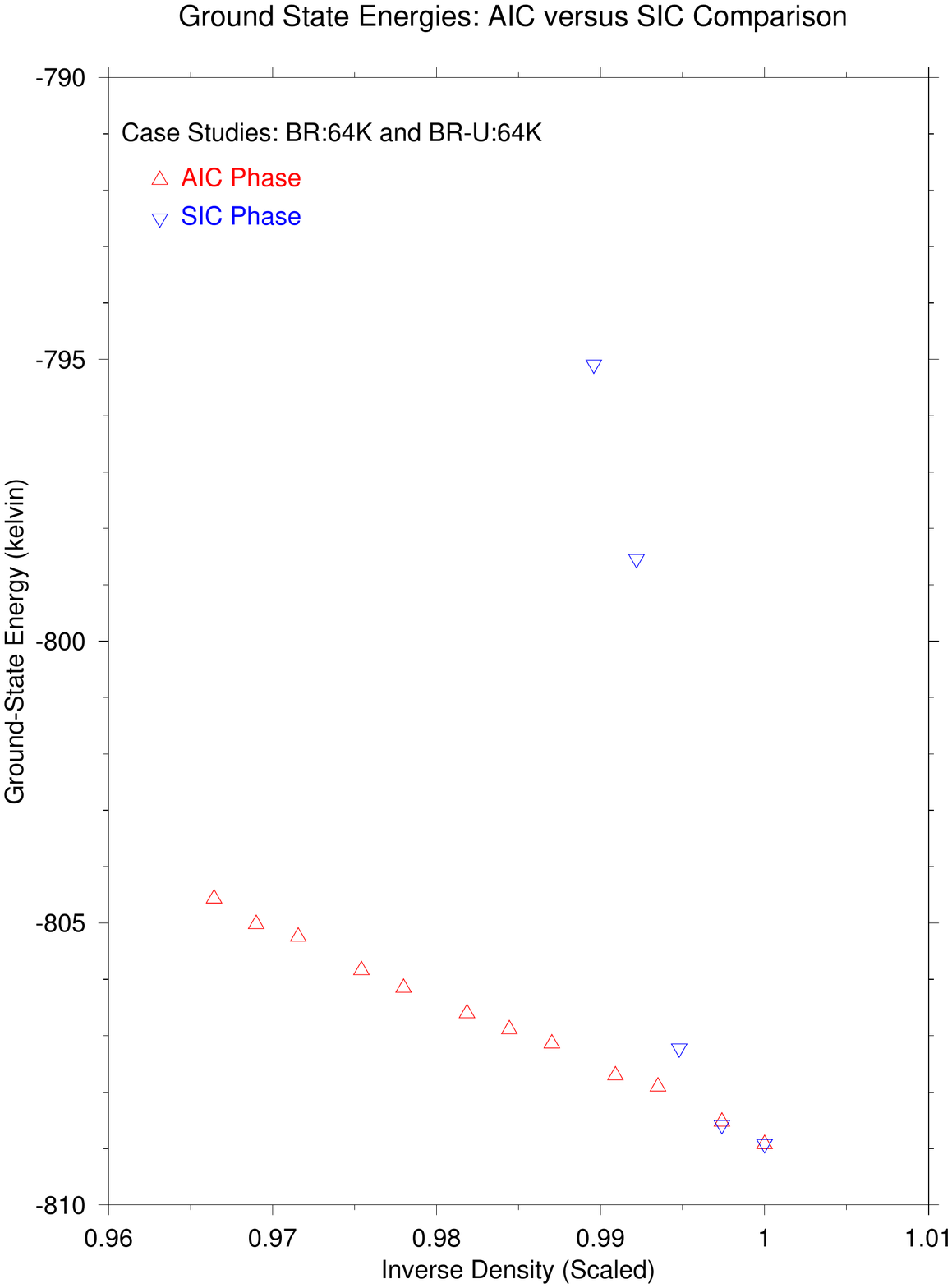}
   \caption{\label{Fig_GSE.vs.InvDens} Ground state energy
    for constrained simulations, comparing the AIC  phase (case study
    BR:64K) and the SIC phase (case study BR-U:64K).  The density is scaled
    to that of the \Rt{3} phase.}
\end{figure}
An examination of the ground state of the constrained geometry was
carried out for the BR projection because it has the more appropriate
corrugation for very low temperatures.  The classical ground-state energy
is obtained by an extrapolation of the finite temperature total energy to
zero temperature.\cite{BruNov08} As an example, comparisons for the BR
projection using a system size of $64$K and initial states of the \Rt{3},
AIC, and SIC structures, show that the stable, zero pressure ground state is
the \Rt{3} phase.  Figure~\ref{Fig_GSE.vs.InvDens} shows a plot of the
ground-state energy of the AIC and SIC phases as a function of inverse density.
It is clear that the \Rt{3} structure is the stable zero pressure phase.
At finite misfits, the AIC phase is decidedly preferred to the SIC phase
except for small misfits (having scaled densities $> 1.00$ but $< 1.003$).
For these small misfits, the difference in energy
between the AIC state and the SIC state is zero within statistical uncertainty.
It is reasonable to assume that these small differences between the energies of
the AIC and SIC phases could be an important driver for the
``chaotic'' behavior of the domain structure that
is observed in this system.  Similar results are found in the unconstrained
case studies as shown in Sec.~\ref{SuplUnconstrained} below.

Preliminary calculations of quantum effects (using a SCP approach for large
clusters) indicate that these conclusions are not substantially altered by such
effects and that the basic behavior described here also holds for low
temperatures when quantum effects are included.\cite{Note-Novaco}
The preliminary calculations show free energy curves that look much like
the energy curves in Fig.~\ref{Fig_GSE.vs.InvDens}.

\subsection{\label{SuplUnconstrained}Unconstrained Geometry}

The initialized structures appear to be, after some initial
relaxation into a quiescent state, quite stable.  However, different
initializations often stabilized with different structures and even
different energies existing at the same system temperature.
This is taken to be an indication that the system settles into meta-stable
states which have long lifetimes on the scale of the MD simulation.
Since the stable state at finite temperature is determined by the lowest free
energy, the arguments in Sec.~\ref{FEdiscuss} are
used to determine the most likely stable state at finite temperatures.
\begin{figure}
   \centering
   \includegraphics[width=2.8in]{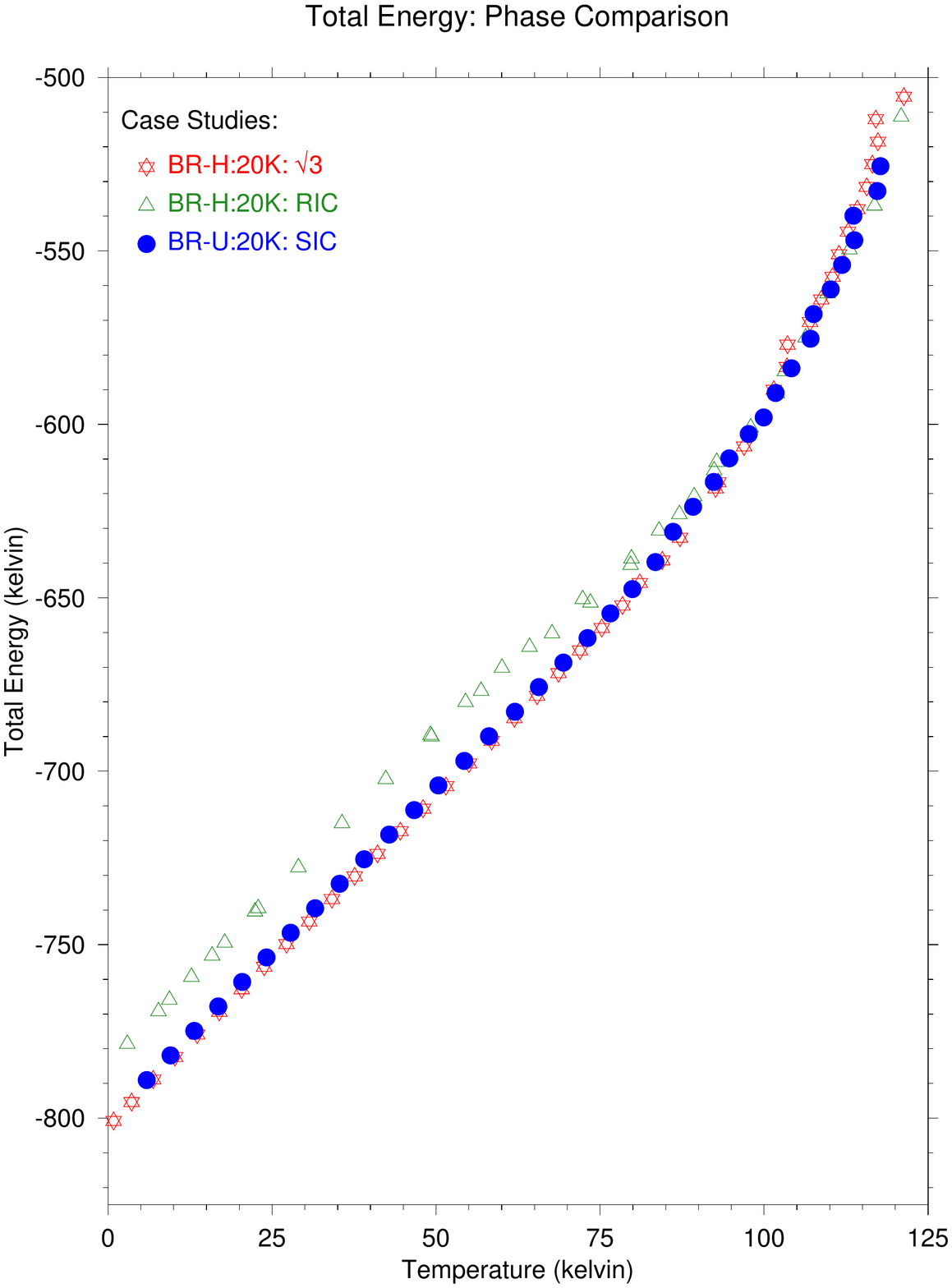}
   \caption{\label{Fig_TotEng.vs.Temp_Phases} The total energy
    of the case studies for BR-X:20K, comparing initializations in the \Rt{3},
    RIC, and SIC initial states for the unconstrained geometry.
    They all have essentially the same average density of 0.5$\rho_0$, and
    only the heating data is shown. }
\end{figure}

Figure~\ref{Fig_TotEng.vs.Temp_Phases} compares the total energy of
unconstrained \Rt{3} and corresponding meta-stable RIC and SIC initializations
for the BR-X:20K case study.
The \Rt{3} initialization shown starts with a hexagonal patch having
a single domain and minor imperfections on the boundary of the patch.  The
RIC initialization starts with a hexagonal patch of the \Rt{3} initialization,
but then rotates it by $2^{\circ}$ counterclockwise before starting the
simulation.  The SIC initialization shown starts with a rectangular
patch having almost equal sides and 8 stripes (which are roughly
parallel to the $[1,1]$ direction of the xenon lattice).  It can be seen
that the \Rt{3} points are lower in energy than either of the IC data sets.
However, note the closeness of the SIC and the \Rt{3} data sets, even though
the SIC structure was initialized with the maximum number of stripes
possible (8) without mechanical instabilities causing the simulation
to abort.

At some temperature below
$100$~K, the three separate data point sets merge into a single set of points
exhibiting a \Rt{3} like structure, thus presumably satisfying the conditions
set forth in Sec.~\ref{FEdiscuss}. Using the
free energy analysis of that section, it appears that the \Rt{3} phase is,
in this temperature range, the most stable state of the three.
However, the SIC phase is very close to the \Rt{3} phase in (free) energy,
implying that the SIC can easily compete with the \Rt{3} phase.
The MD results do imply that not only is the \Rt{3} phase the ground
state of this system, it is always stable against both the RIC and SIC phases
until it disorders (melts).\cite{NovBruBav15} The existence of a stable
\Rt{3} phase up to the disordering (melting) transition also occurs for
corrugations more appropriate to the higher temperatures in this figure.
However, the temperature of the transition from the chaotic state to the
\Rt{3} phase does depend upon the value of the corrugation.\cite{NovBruBav15}

\begin{figure}
   \centering
   \includegraphics[width=2.8in]{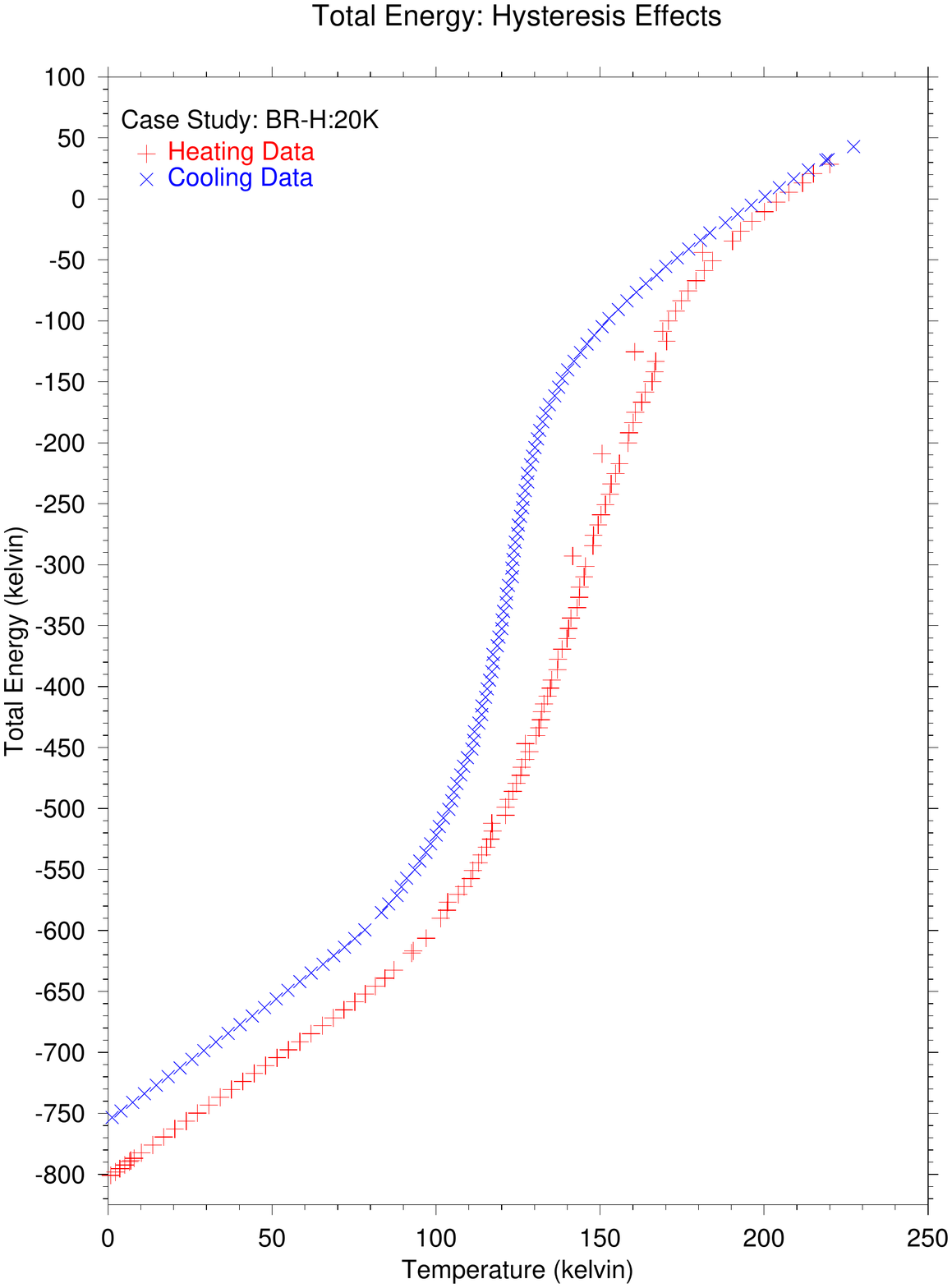}
   \caption{\label{Hysteresis:Energy.vs.Temperature} The
    total energy for case study BR-H:20K using the \Rt{3} initialization.
    Both heating-cycle data (lower curve) and cooling-cycle data
    (upper curve) are shown. Data points off the curve for
    the heating cycle data show the effects of averaging over the time
    interval $[32.6,35.9]$~ns instead of the usual $[13.0,16.3]$~ns.}
\end{figure}

Stability comparisons between different initializations
were carried out only for the heating-cycle data set because
the cooling-cycle data exhibits hysteresis if the system is cooled
from a disordered state.  For example, consider a \Rt{3} initialization
started at a very low temperature, then slowly heated past the
transition to the disordered state, then cooled down to very low temperatures
(close to the initial temperature).
Figure~\ref{Hysteresis:Energy.vs.Temperature}
demonstrates this hysteresis in the case study BR-H:20K,
showing the hysteresis effects of heating and cooling in the simulations. The
upper (cooling cycle) branch, which is unstable to the lower (heating cycle)
one, is associated with a collection of small patches of \Rt{3} structures
instead of the one large patch found in the lower branch.

In the heating process, most runs were carried
out to 16.3~ns, the averages then taken over the last 3.26~ns.  For
a select set of temperatures, the runs were carried out to 35.9~ns, and
the averages again taken over the last 3.26~ns.  Note that the longer runs
shown in Fig.~\ref{Hysteresis:Energy.vs.Temperature} are
the ones slightly to the left of the main heating data set, showing that some
relaxation is still present in the transition from order to disorder.
The cooling data set does not retrace the heating data set, the low
temperature state (upon cooling) being a meta-stable structure with
``islands'' of registered structures with about $100$--$300$ atoms per
island.\cite{NovBruBav15} Using the arguments of
Sec.~\ref{FEdiscuss}, it is clear that the lower data set is the more stable
branch and the upper branch is less so. This is consistent with
the usual arguments used at very low temperatures where the energy term
dominates the free energy.

Assuming the validity of the BR Model, either the BR projection
(strictly classical) or the U30 projection (quantum corrections) would be
the appropriate corrugation for
very low temperatures. Nevertheless, simulations were carried out for other
corrugations at these low temperatures in order to test the stability of the
\Rt{3} ground state to variations in the corrugation.
For $U_{(10)}$ values of $-25$, $-20$, and $-15$ kelvin, all MD simulations
show the IC structure to be the ground state of the system, but still having
a stable \Rt{3} structure at higher temperatures (but still below melting).
If the corrugation is lowered much below $U_{(10)} = -15$~K, there is no
stable \Rt{3} phase at any temperature, the exact value at which this occurs
not being examined here.

The MD specific heat shows no clear sign of a transition near $60$~K.  If the
system is initialized in the \Rt{3} structure at low temperature, the 
$\psi_0$ order
parameter decreases monotonically with increasing temperature, showing smooth
behavior from low temperatures to melting.  The system stays in the \Rt{3}
phase until the disordering temperature is reached.  However, if the system is
initialized in an IC structure at low temperature (thus having a small
$\psi_0$ value), above $60$~K there is an increase in $\psi_0$
with temperature.\cite{NovBruBav15} Some experimental papers\cite{Ker87} claim
a transition from a low temperature striped phase to the \Rt{3} phase as the
temperature is raised above $60$~K. For all the corrugations and
initializations investigated here, the simulations show a gradual transition
from the initial low temperature structure into the \Rt{3} phase at higher
temperatures, but the temperature range of this transition does depend upon
both the corrugation and the initialization. The transition
does not seem to be a true
thermodynamic transition, but rather appears to be a gradual evolution into
the \Rt{3} structure.  However, this apparent behavior could be a consequence
of the very long relaxation times for this system and not a reflection of the
system's true equilibrium thermodynamics.

\subsection{\label{SuplStructure}Structure of the Monolayer}

Our simulation shows there is no tendency for the sub-monolayer patch to
spontaneously rotate away from the
high symmetry direction of the substrate.  This behavior is in contrast to
that of Xe/Gr,\cite{NovBru14} and it is probably due to Xe/Pt having
a much stronger corrugation and a dilated lattice
compared to Xe/Gr.  Even when the issue is
forced by the generation of a rotation in the initial state,
the imaginary part of $\psi_6$ quickly becomes very small
(about $10^{-2}$ once the system has stabilized) thus indicating a near zero
rotation angle.\cite{NovBru14}  A small initial rotation angle generates
rather irregular domain shapes, the typical size of these domains being
dependent upon that angle.  However, unlike the Xe/Gr case where the whole
patch rotates to the stable angle,\cite{NovBru14} the edges of these
Xe/Pt patches remained (approximately) at the original rotated orientation with
respect to the substrate.  The adatoms in the domains simply re-position
themselves to produce local alignment with the \Rt{3} state orientation, and
the domain walls respond (relax) accordingly, forming a
chaotic, meta-stable structure.

Initial structures generated by
small initial rotations result in large domains of the \Rt{3} structure
separated by narrow walls (generally zero to 2 atoms wide) that run in a
irregular manner through the system.  Those generated by large rotation
angles tend to have smaller domains possessing more regular boundaries.
Using the analysis outlined in Sec.~\ref{FEdiscuss},
these various structures are found to be thermodynamically unstable
to the \Rt{3} state.

Initializations generated by small increases in the initial density tend to
have larger domains with domain walls that tend to be straighter but otherwise
similar to the walls in the initializations generated by a rotation.
The range of interior patch densities that can be used to initialize an
unconstrained geometry without the system becoming mechanically unstable is
quite small.  While there is no simple and reliable way to estimate the
patch
density, the densest initialization had about 6~\% of the atoms in irregular,
super-heavy-like domain walls with the others in the three types of
domains\cite{NovBru14} having domain populations of roughly 25 to 50~\%.
Attempts to initialize with higher patch densities result in simulations
aborting with loss of energy conservation.  As stated previously, the \Rt{3}
patch is the thermodynamically stable state, but the others can be
thermodynamically meta-stable, often without any major reconstructions over
16.3~ns or longer.  Only selected runs were tested beyond this, but they too
were meta-stable.

The initial relaxation of the system often results in domains of various
geometries without a clear pattern.  When the simulation was allowed to evolve
at fixed energy, the walls seem to be pinned down and rather resistant to
movement.  This is true even when thermodynamic analysis indicates that the
state in question is not a  stable structure.  Rather, these structures do
appear to be meta-stable on the time scale of any reasonable MD simulation.
We label such structures ``chaotic''.
This is in marked contrast to the behavior of the walls
found in the Xe/Gr system, where the domain walls have a regular structure,
are relatively wide, and are of essentially constant
width.\cite{BruNov08,NovBru14} Furthermore, the walls in the Xe/Pt system,
at least for the unconstrained cases, do not match the description of domain
walls used  in many, if not most, of the calculations in the literature on
this subject.\cite{KerDavZep87,SpeMakPet87,KarBer82}

\section{\label{SuplDiscusion}Existence of Metastability}

Our data shows strong evidence of meta-stability and hysteresis. However,
the difficulty of making statements about meta-stability based upon
the results of any MD simulation is two-fold. First: The time scale of any
practical MD simulation is very small on the scale of macroscopic experiments,
the longest run in these simulations corresponding to about 49~ns.
Second: The system size of the simulation cell is usually much smaller
than the macroscopic system.  Although the justification here is a bit sounder,
as the largest simulation presented here is comparable to the smaller
experimental examples, the size is still smaller than the best
experimental cases.  Nevertheless, even with those caveats, it is clear
that the meta-stability observed here is qualitatively different than that
observed for Xe/Gr simulation.\cite{BruNov08,NovBru14}
While it is not possible to make a definitive conclusion,
one can still draw the reasonable inference that meta-stability must
be considered a strong possibility in this system.  This thought is bolstered
by the fact that the free energy analysis of the MD results shows that, while
the \Rt{3} phase is the stable phase for the BR model (the unconstrained
geometry case) below the disordering temperature, the SIC phase (with a small
misfit) is very close in free energy to
the \Rt{3} phase.  It is then possible that any lack of thermal equilibrium
could drive competition between these (and possibly other) states.  In fact,
we have observed  ``chaotic'' states which consisted of broad bands of \Rt{3}
domains, which were much longer than they were wide, mixed in with
configurations that have a roughly hexagonal shape.  Figure 5 in the
main text is one such example.  These meta-stable structures
are interesting, not only in their own right, but because of their relation
to the general study of chaotic systems and even, perhaps, to our
understanding of the third law of thermodynamics.\cite{MasOpp17,AizLie81}

%\bibliography{Citations,Books-Cite,FootNotes}
%\bibliographystyle{apsrev}

%\twocolumngrid